\documentclass[twocolumn,aps,preprint,amssymb,superscriptaddress,10pt]{revtex4}
\usepackage{graphicx}
\usepackage{subfigure}
\newcommand\etal{{\it et al.}}

\begin{document}

\title{Nuclear ``pasta'' structures in low-density nuclear matter and neutron star crust}

\author{Minoru Okamoto}
\affiliation{Graduate School of Pure and Applied Science, University of Tsukuba, Tennoudai 1-1-1, Tsukuba, Ibaraki 305-8571, Japan}
\affiliation{Advanced Science Research Center, Japan Atomic Energy Agency, Shirakata Shirane 2-4, Tokai, Ibaraki 319-1195, Japan}
\author{Toshiki Maruyama}
\affiliation{Advanced Science Research Center, Japan Atomic Energy Agency, Shirakata Shirane 2-4, Tokai, Ibaraki 319-1195, Japan}
\author{Kazuhiro Yabana}
\affiliation{Center of Computational Sciences, University of Tsukuba, Tennoudai 1-1-1, Tsukuba, Ibaraki 305-8571, Japan}
\affiliation{Graduate School of Pure and Applied Science, University of Tsukuba, Tennoudai 1-1-1, Tsukuba, Ibaraki 305-8571, Japan}
\author{Toshitaka Tatsumi}
\affiliation{Department of Physics, Kyoto University, Kyoto 606-8502, Japan}

\begin{abstract}
 In neutron star crust, non-uniform structure of nuclear matter is expected, which is called the ``pasta'' structure. 
 From the recent studies of giant flares in magnetars, these structures might be related to some observables and physical quantities of the neutron star crust. 
 To investigate the above quantities, we numerically explore the pasta structures with a fully three-dimensional geometry and study the 
 properties of low-density nuclear matter, based on the relativistic mean-field model and the 
 Thomas-Fermi approximation. 
 We observe typical pasta structures for fixed proton number-fraction and two of them for cold catalyzed matter. 
 We also discuss the crystalline configuration of ``pasta''.
\end{abstract}

\maketitle

\section{Introduction}
A neutron star has a radius of about 10 km and the mass about 1.4 times
the solar mass. It is considered to consist of four parts \cite{page,haensel}. 
The region around 0.3km from the surface is called ``outer crust'',
where Fe nuclei are expected to form a Coulomb lattice. 
The region around 0.3--1 km is called ``inner crust'' with a density about 0.3--0.5 $\rho_0$. 
There are neutron-rich nuclei in lattice and dripped neutrons in a superfluid state. 
Two central regions with higher densities are called ``outer core'' and ``inner core'', where speculated 
are proton superconductivity, neutron superfluidity \cite{takatsuka}, meson condensations \cite{pion1,pion2,mesoncond,prakash}, 
hyperon mixture \cite{hypmix,glend-hyp,ishizuka,tsubakihara}, or quark matter \cite{quark1,quark2}.
The transition of matter composition with the change of density inside neutron stars causes a question:
does it change smoothly or suddenly? A sudden change of matter property is generally accompanied
by a first-order phase transition which causes an appearance of the mixed phase.
Ravenhall \etal \cite{ravenhall} suggested the existence of non-uniform structures of nuclear matter, i.e. the structured
mixed phase. They suggested five types of geometrical structures such as droplet, rod, slab, tube, and bubble
\footnote{Note that emergence of the inhomogeneous structure is a general feature accompanying the non-congruent first order phase 
transition \cite{NSbook,hempel13}, and there have been known other examples including the hadron-quark deconfinement or meson condensation \cite{maruHQ,yasutake}.}. 
Due to its geometrical shapes which depend on the density, we call it ``nuclear pasta'' like spaghetti and lasagna
etc \cite{ravenhall,hashimoto}. 
Many studies have suggested the existence of the pasta structures in low-density nuclear matter,
relevant to the crust region of neutron stars and the collapsing stage of supernovae. 
The existence of the
pasta structures in the crust of neutron stars may not have influence on the bulk property and structure of
neutron stars so much. However, it should be important for the mechanism of glitch, the cooling process of neutron
stars, and the thermal and mechanical properties of supernova matter \cite{mochizuki}.

Recently, in the X-ray afterglow of giant flares 
quasi-periodic oscillations (QPOs) have been observed in some soft-gamma ray repeaters (SGR) \cite{tod,watts}. 
These flares are energetic $\gamma$-ray bursts from strongly magnetized neutron stars, magnetars. 
As one comprehensible understanding of the QPOs they may be attributed to the shear oscillations of neutron star crust. 
Based on this interpretation, detailed information on neutron star crust can be extracted from the QPOs \cite{sotani}, where  
the frequencies of shear oscillations depend on shear moduli of the Coulomb lattice of nuclei. 
Ogata \etal\ have calculated shear moduli using molecular dynamics simulations \cite{ogata}. 
However, in this calculation, some important effects are missing, 
such as charge screening, finite-size effects of nuclei, superfluidity of dripped neutrons and so on \cite{chamel}. 
Considering the realistic situation, these elements should be taken into account in the elaborate calculation of shear moduli. 
Pasta structures should be also considered.
Thus the energy change against a small deformation of the lattice can be discussed using shear moduli. 
On the other hand, if the energy change against a large deformation is known, the breaking strain of neutron star crust will be obtained. 
Then, this breaking strain could determine the possible size of mountains on the neutron star crust, which may radiate gravitational waves strongly with rapid rotation \cite{kadau}. 
These waves may be caught by large-scale interferometers and could limit the spin frequencies of accreting stars . 
Furthermore, the breaking strain may be important for the 
``star quake'' model of giant flares in magnetars \cite{horowitz}.

The species and the sizes of the pasta structure 
are determined for given average baryon-number densities by 
minimizing the total energy density, which consists of the bulk, the surface, and the Coulomb energy densities. 
From the thermodynamical point of view, nuclear matter at sub-saturation density can be represented by 
a dilute gas phase and a dense liquid phase in chemical equilibrium which determines particle densities in both phases. 
Once the averaged density is given, the volume fraction of gas or liquid phase is determined. 
Thus, the bulk energy density is independent of the shape and the size of the structure, which 
are determined by the balance 
between the Coulomb and surface energy densities \cite{bbp}.

In many studies about nuclear pasta, the geometrical symmetry of the structure has been assumed 
by employing the Wigner-Seitz (WS) approximation \cite{ravenhall,hashimoto}. 
In this approximation, one can find all the physical quantities from those in a single WS cell. 
Furthermore, the calculation is reduced to  the one-dimensional form owing to the geometrical symmetry, 
which drastically saves the computational cost. 
On the other hand, by virtue of the recent development in computational science, 
it becomes possible to calculate without any assumption about the geometry 
of non-uniform structures
\cite{williams,newton}. 
These studies have imposed the periodic boundary condition in a small cubic cell, which included only one period of structures.  
Although they got essentially the same results, i.e.\ typical pasta structures, 
it is hard to  extract further information such as crystalline configuration of ``pasta'' and mechanical properties.

For detailed studies about the properties of neutron star crust, 
it is desired to perform three-dimensional calculations in a periodic cubic cell with sufficiently large sizes. 
Accordingly, we have developed a numerical code to calculate the density distribution of particles
in a three-dimensional coordinate space and the relevant physical quantities, e.g., energy density or pressure of matter, 
by using a relativistic mean-field (RMF) model under the Thomas-Fermi approximation \cite{okamoto}.
We have explored ground states of low-density nuclear matter with a fixed proton number-fraction and observed a series of typical ``pasta'' structures appearing as the ground states.
We have also observed crystalline configurations of droplets, rods, tubes, and bubbles.
One of our findings is the appearance of face-centered cubic (fcc) lattice of droplets in the ground state.
However, the system was limited to the cases with fixed proton number-fraction 
and detailed discussion is still needed about its appearance. 

In this article, we explore the pasta structures and properties of low-density
nuclear matter, following the line of the previous study in Ref.\ \cite{okamoto}
but in more detail and with wider scope to include catalyzed matter.
Appearance of the pasta structures and other unusual structures, crystalline configuration of droplets are discussed.

In Sec.\ II, we present the model and describe our numerical procedure. 
In Sec.\ III, we first demonstrate some results from our three-dimensional calculation for 
low-density nuclear matter with fixed proton number-fraction $Y_p=$ 0.5, 0.3, 0.1 that may 
be related to the supernova matter and newly born hot protoneutron star crust. 
Then, in the second part of Sec.\ III, we investigate the pasta structures at $\beta$-equilibrium, 
as they occur in cold neutron stars, and the crystalline configuration of nuclear pasta. 
Finally, Sec.\ IV is devoted for summary and concluding remarks.

\section{Model and Method}

There  have been used the many-body techniques for studying the pasta structures in the literature, 
such as compressible liquid-drop model (CLDM) \cite{ravenhall,hashimoto,oyamatsu,nakazato,matsuzaki}, 
the Thomas-Fermi model\cite{williams,ogasawara,boguta,avancini,maruyama}, 
the Hartree-Fock (HF) approximation employing effective NN interactions \cite{newton,iwata,schuetrumpf}, and quantum molecular dynamics (QMD) \cite{qmd_super,qmd1,qmd2}. 
In the studies using CLDM and the Thomas-Fermi model, always used was the WS approximation, 
where only the typical pasta structures are assumed. 
QMD calculation does not assume any specific non-uniform structure of baryon matter, while uniform background of electrons, though their distributions should be almost uniform, is assumed. 
Some of the Thomas-Fermi calculations and the HF calculation used a periodic boundary condition and did not assume 
any geometrical symmetry for the structure.
However, the size of the periodic unit cell was not large enough for quantitative discussion. 
In this paper, to describe interaction among nucleons, we employ an RMF model under the Thomas-Fermi approximation \cite{maruyama}. 
It deals with the mean-fields of sigma, omega, and rho mesons represented by $\sigma,\omega^\mu$, and ${\bf R}^\mu$, respectively, together with nucleons $\psi$, electrons $\psi_e$, and electromagnetic field $A_\mu$, by a Lagrangian $\cal L$ introduced in a Lorentz-invariant form as follows, 

\begin{eqnarray}
 \nonumber
 {\cal L} &=& {\bar{\psi}} [ i{\gamma}^{\mu}{\partial}_{\mu}
                         - m_{N}^{*}- g_{{\omega}N}{\gamma}^{\mu}{\omega}_{\mu} 
			 - g_{{\rho}N}{\gamma}^{\mu}{\mbox{\boldmath$\tau$}}{\cdot}{\mbox{\boldmath$R$}}_{\mu}
\\
 \nonumber
    & &                              -e{\frac{1+{\tau}_3}{2}}{\gamma}^{\mu}A_{\mu} ]{\psi} 
 \nonumber
    + {\frac{1}{2}}({\partial}_{\mu}{\sigma})^2 - {\frac{1}{2}}m_{\sigma}^2{\sigma}^2 - U({\sigma}) \\
 \nonumber 
    & & -{\frac{1}{4}}{\omega}_{\mu \nu}{\omega}^{\mu \nu} + {\frac{1}{2}}m_{\omega}^2 {\omega}_{\mu}{\omega}^{\mu}
     - {\frac{1}{4}}{\mbox{\boldmath$R$}}_{\mu \nu}{\mbox{\boldmath$R$}}^{\mu \nu}
     + {\frac{1}{2}}m_{\rho}^2{\mbox{\boldmath$R$}}_{\mu}{\mbox{\boldmath$R$}}^{\mu} \\
    & & - {\frac{1}{4}}F_{\mu \nu}F^{\mu \nu}
     + {\bar{{\psi}_e}}{\left [ i{\gamma}^{\mu}{\partial}_{\mu} - m_e + e{\gamma}^{\mu}A_{\mu} \right]}{\psi}_e, 
\label{Lagrangian}
\end{eqnarray}
where
$U(\sigma)= \frac{1}{3} bm_N (g_{{\sigma}N}{\sigma})^3 - \frac{1}{4} c(g_{{\sigma}N}{\sigma})^4$
is a non-linear term for the scalar field,
$m_N^*({\bf r})=m_N-g_{\sigma N}\sigma({\bf r})$ represents an effective mass of nucleon, and $g_{ab}$ represent coupling constant between $a$ and $b$. Thermodynamic potential $\Omega$ is then given by $\Omega=E-\sum_{i=n,p,e}\mu_i\int d^3r\rho_i({\bf r})$ in terms of total energy $E$, number densities  $\rho_i({\bf r})$ and chemical potentials $\mu_i$. Variation of $\Omega$ with respect to each field gives the field equation.

The set of coupled field equations for the mean-fields 
and the Coulomb potential $A^\mu=(V_{\rm Coul},{\bf 0})$, render  
\begin{eqnarray}
&-&{\nabla}^{2}{\sigma}({\bf r})+m^{2}_{\sigma}{\sigma}({\bf r}) =
 g_{{\sigma}N}\left(\rho_p^s({\bf r})+\rho_n^s({\bf r})\right)
 -\frac {dU}{d{\sigma}} \label{sigma_rmf},\\
&-&{\nabla}^{2}{\omega}_{0}({\bf r})+m_{\omega}^{2}\omega_{0}({\bf r}) =
 g_{{\omega}N}(\rho_{p}({\bf r})+\rho_{n}({\bf r}))\label{omega_rmf},\\
&-&{\nabla}^{2}R_{0}({\bf r})+m_{\rho}^{2}R_{0}({\bf r})=
  g_{{\rho}N}({\rho}_{p}({\bf r})-{\rho}_{n}({\bf r}))\label{rho_rmf},\\
& &{\nabla}^{2}V_{\rm Coul}({\bf r}) = 4\pi e^2\left({\rho}_p({\bf r}) - {\rho}_e({\bf r})\right),
 \label{Coulomb_eq}
\end{eqnarray}
where $\rho_i^s({\bf r})=\langle \bar{\psi}_i({\bf r}){\psi}_i({\bf r}){\rangle}, i=p,n$ is the nucleon scalar density
and $R_0$ is the third component of the isovector field ${\bf R_\mu}$.
Field equations for fermions simply yield the standard relations between the densities $\rho_i$ and chemical potentials $\mu_i$ within 
the Thomas-Fermi approximation,
\begin{eqnarray}
\mu_n 
       &=& \sqrt{k_{{\rm F},n}({\bf r})^2+{m_N^*({\bf r})}^2}
         +g_{\omega N}\omega_0({\bf r})-g_{\rho N}R_0({\bf r}) \label{eq:cpotB}\nonumber\\\\
\mu_p 
      &=& \sqrt{k_{{\rm F},p}({\bf r})^2+{m_N^*({\bf r})}^2}+g_{\omega N}\omega_0({\bf r})+g_{\rho N}R_0({\bf r})\nonumber\\ 
      & & \hspace{50mm} -{V_{\rm Coul}({\bf r})} \label{eq:cpotBp} \\
{\rho_e({\bf r})}&=& (\mu_e-{V_{\rm Coul}({\bf r})})^3/3\pi^2, \label{eq:rhoe}
\end{eqnarray}
where the local Fermi momentum, $k_{F,i}({\bf r})$, is simply related to the density, $k^3_{F,i}({\bf r})/(3\pi^2)=\rho_i({\bf r})$.
Finally the baryon-number conservation and the charge neutrality are imposed besides these equations.
We use the same set of parameters as in Ref.\ \cite{maruyama} listed in Table\ \ref{parameter}, 
in order to compare the equation of state (EOS) and structural changes of the pasta structure with and without the WS approximation. 
With these parameters, we can reproduce the properties of uniform nuclear matter shown in Table \ref{EOS_table}. 
The first and second quantities, $\rho_0$ and $\epsilon_0$, are the saturation density of symmetric nuclear matter 
($\approx 0.16\ \rm fm^{-3}$) and its energy per nucleon, respectively.
The third and forth quantities $K$ and $S_0$ are the incompressibility and symmetry energy at $\rho_0$, respectively.
The last one $L$ is  the  slope parameter of  symmetry energy at $\rho_0$.
Using these parameters the binding energy per nucleon around the saturation density is expressed as
\begin{eqnarray}
	\frac{E}{A} &=& \epsilon_0+\frac{K(\rho-\rho_0)^2}{18\rho_0^2}+
	\left[S_0+\frac{L(\rho-\rho_0)}{3\rho_0}\right](1-2Y_p)^2.\nonumber\\
\end{eqnarray}

\begin{table}[htbp]
\begin{center}
\begin{ruledtabular}
 \begin{tabular}{cccccc}
	 $\rho_0$ [fm$^{-3}$] &  $\epsilon_0$ [MeV] &  $K$ [MeV] &  $S_0$ [MeV] &  $L$ [MeV] \\
  \hline
  0.153  &    $-16.4$    &   240      &   33.4    & 84 \\
 \end{tabular}
\end{ruledtabular}
 \caption{EOS of uniform nuclear matter.}
 \label{EOS_table}
\end{center}
\end{table}

\begin{table*}[htbp]
\begin{center}
\begin{ruledtabular}
 \begin{tabular}{ccccccccc}
  $g_{\sigma N}$ & $g_{\omega N}$ & $g_{\rho_N}$ & $b$         & $c$           &
   $m_{\sigma}$ [MeV] & $m_{\omega}$ [MeV] & $m_{\rho}$ [MeV]  \\
  \hline
  6.3935       & 8.7207      & 4.2696    & 0.008659 & $-0.002421$ &
  400                  &  783                & 769     \\
 \end{tabular}
\end{ruledtabular}
 \caption{Parameter set used in RMF model.}
 \label{parameter}
\end{center}
\end{table*}

To numerically simulate the non-uniform structure of infinite matter, we use a cubic cell with a periodic boundary condition. 
If the cell size is small that includes only one or two units of the structure, the geometrical shape should be affected by the boundary condition and the appearance of some structures is implicitly suppressed. 
Therefore, the cell size should be so large as to include several units of the pasta structures. 
We divide each cell into three-dimensional grids. 
The desirable grid width should be so small as to describe the detailed density distribution,  particularly at the nuclear surface. 
Due to this requirement, we set the grid width to 0.3 fm at the largest. 

\begin{figure}
\begin{center}
 \includegraphics[width=0.23\textwidth]{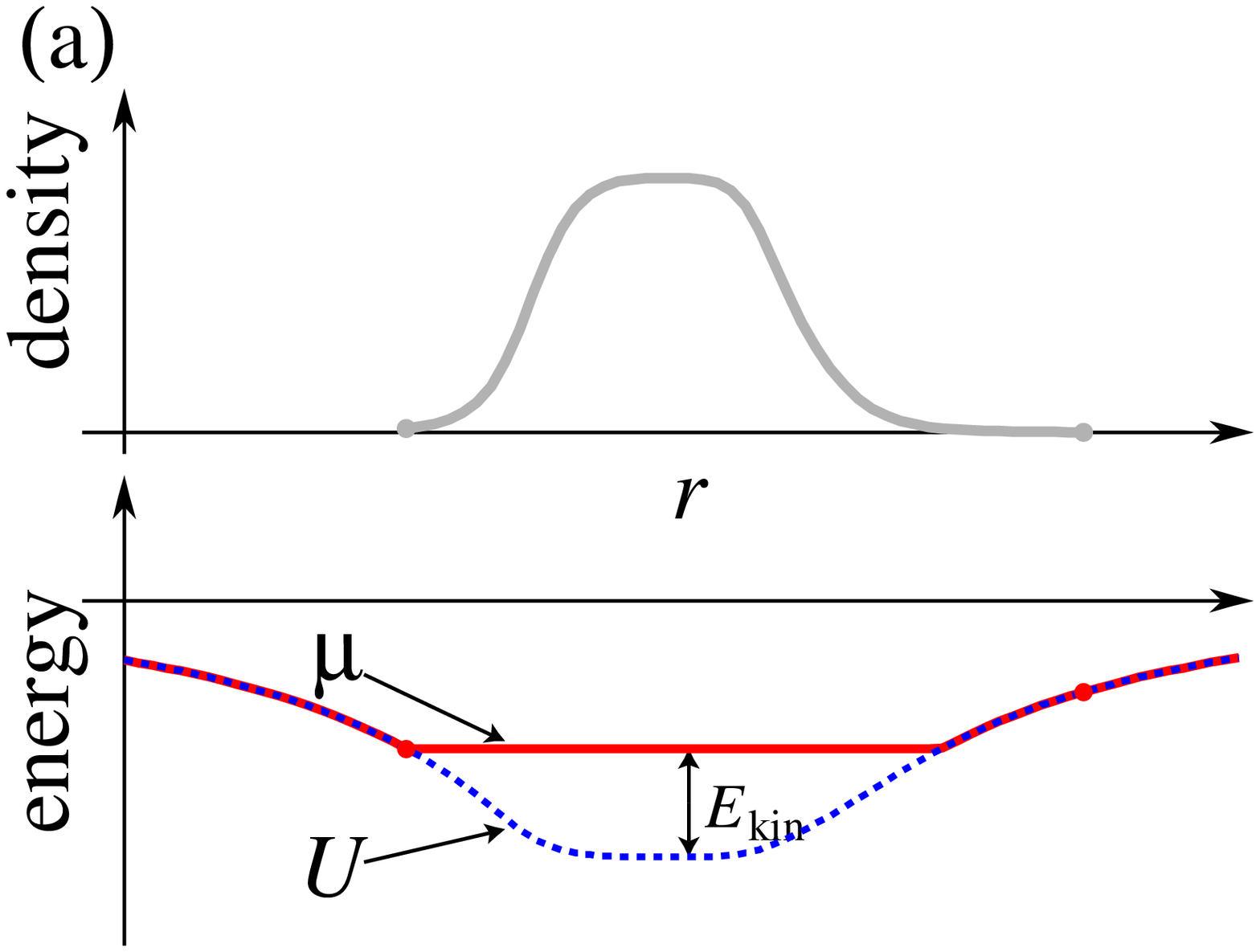}
 \includegraphics[width=0.23\textwidth]{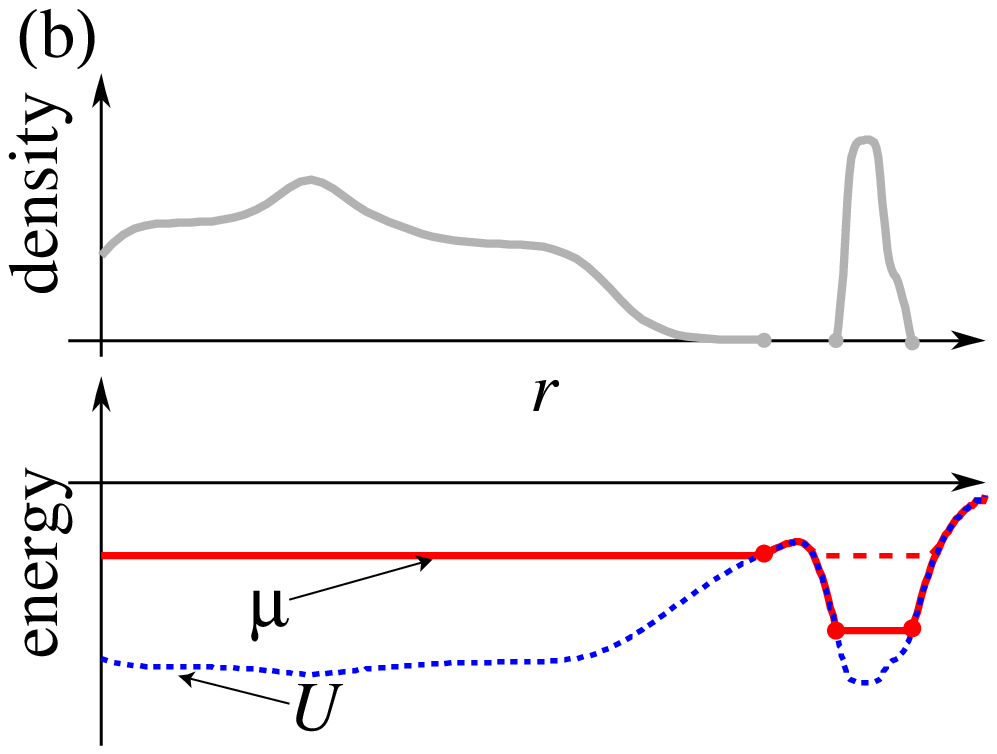}
 \caption{
(a)
Schematic figure of density distribution and the chemical potential subtracting 
a constant mass.
In the region with zero density, where the kinetic energy is zero,
the chemical potential $\mu({\bf r})$ is identical to the potential $U({\bf r})$,
while the chemical potential in the region with finite particle density is constant.
\\
(b)
Schematic figure of non-uniform matter with an isolated matter region, 
where the chemical potential is different from that of the global one.
By the procedure to adjust densities between only the neighboring grids,
such unphysical region of isolated matter may appear.
}\label{figCpot}
\end{center}
\end{figure}

Giving the average baryon-number  density $\rho_B$, initial density distributions of fermions are randomly prepared on each grid point. 
Then proper density distributions and the meson mean-fields are searched for. 
We introduce the local chemical potentials $\mu_a({\bf r})$ ($a=p,n,e$) to obtain the density distributions of baryons and electrons,. 
The equilibrium state is eventually determined so that the chemical potentials are independent of the position. 
An exception is the region with no particle density, where the chemical potential of that particle can be higher 
(see Fig.\ \ref{figCpot}(a) for more explanation). 
We repeat the following procedures to attain uniformity of the chemical potentials. 
A chemical potential $\mu_i({\bf r})$ of a baryon $i=p,n$ on a grid point ${\bf r}$ is compared with those on the 
six neighboring grids ${\bf r}'={\bf r}+d{\bf r}$, ($d{\bf r}=\pm d{\bf x},\pm d{\bf y},\pm d{\bf z}$). 
If the chemical potential at the point under consideration is larger than that of another $\mu_i({\bf r})>\mu_i({\bf r}')$, 
some part of the density will be transferred to the other grid point. 
This adjustment of the density distribution is simultaneously done on all the grid points. 
In addition to the above process, we adjust the particle densities between distant grid points chosen randomly 
in order to eliminate regions with different $\mu_i$ 
which can happen to  isolated matter regions as Fig.\ \ref{figCpot}(b). 
The meson mean-fields and the Coulomb potential are obtained by solving Eqs.\ ({\ref{sigma_rmf})-(\ref{Coulomb_eq}) 
using the baryon density distributions $\rho_i({\bf r})$ ($i=p,n$) 
and the charge density distribution $\rho_p({\bf r})-\rho_e({\bf r})$. 
The electron density $\rho_e({\bf r})$ is directly calculated from 
the Coulomb potential $V_{\rm Coul}({\bf r})$ and  the electron chemical potential $\mu_e$, $\rho_e({\bf r})=(\mu_e-V_{\rm Coul})^3/(3\pi^2)$. 
The global charge neutrality is then achieved by adjusting $\mu_e$. 
Above processes are repeated many times until we get convergence.

We have used PRIMERGY BX900 of JAEA massively parallel computing system. 
To complete a calculation of a typical case, about 510 CPU hours is needed at least.

\section{Results}
\subsection{Fixed proton number-fraction}
First, we present here some results for fixed proton number-fraction $Y_p$ with $Y_p$=0.5 (symmetric nuclear matter), 0.3, and 0.1
which are roughly relevant to the supernova and the neutron star crust. 
Shown in Fig.\ \ref{pasta_fixed} are the proton density distributions in cold symmetric matter. 
We can see that the typical pasta phases with rod, slab, tube, and bubble, in addition to spherical nuclei (droplet), 
are reproduced by our calculation in which no assumption on the structures was used. 
Furthermore, these cells include several units and we can specify these lattice structures. 
Crystalline configuration of droplet and bubble is ``fcc'', rod and tube is ``honeycomb''.

\begin{figure}[htbp]
 \begin{center}
  \includegraphics[clip,width=0.5\textwidth]{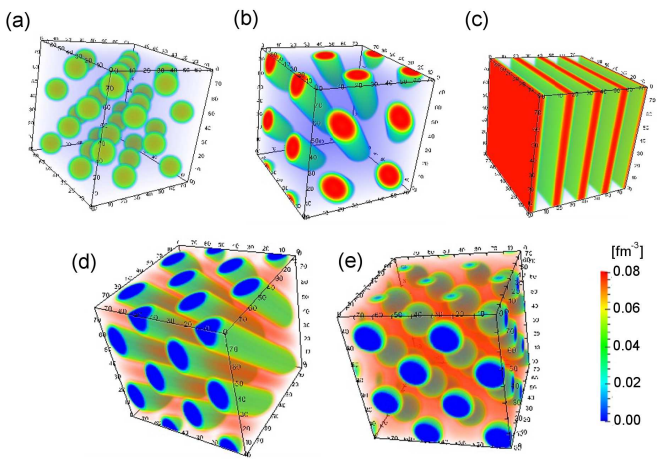}
  \caption{(color online) Proton density distributions in the ground states of symmetric matter ($Y_p=$0.5).
  Typical pasta phases are observed: 
  (a) Spherical droplets with a fcc crystalline configuration at baryon density $\rho_B= 0.01$ fm$^{-3}$.
  (b) Cylindrical rods with a honeycomb crystalline configuration at 0.024  fm$^{-3}$.
  (c) Slabs at 0.05 fm$^{-3}$.
  (d) Cylindrical tubes with a honeycomb crystalline configuration at 0.08  fm$^{-3}$.
  (e) Spherical bubbles with a fcc crystalline configuration at 0.09  fm$^{-3}$.}
  \label{pasta_fixed}
 \end{center}
\end{figure}

\begin{figure}[htbp]
 \begin{center}
  \subfigure{\includegraphics[width=0.6\linewidth]{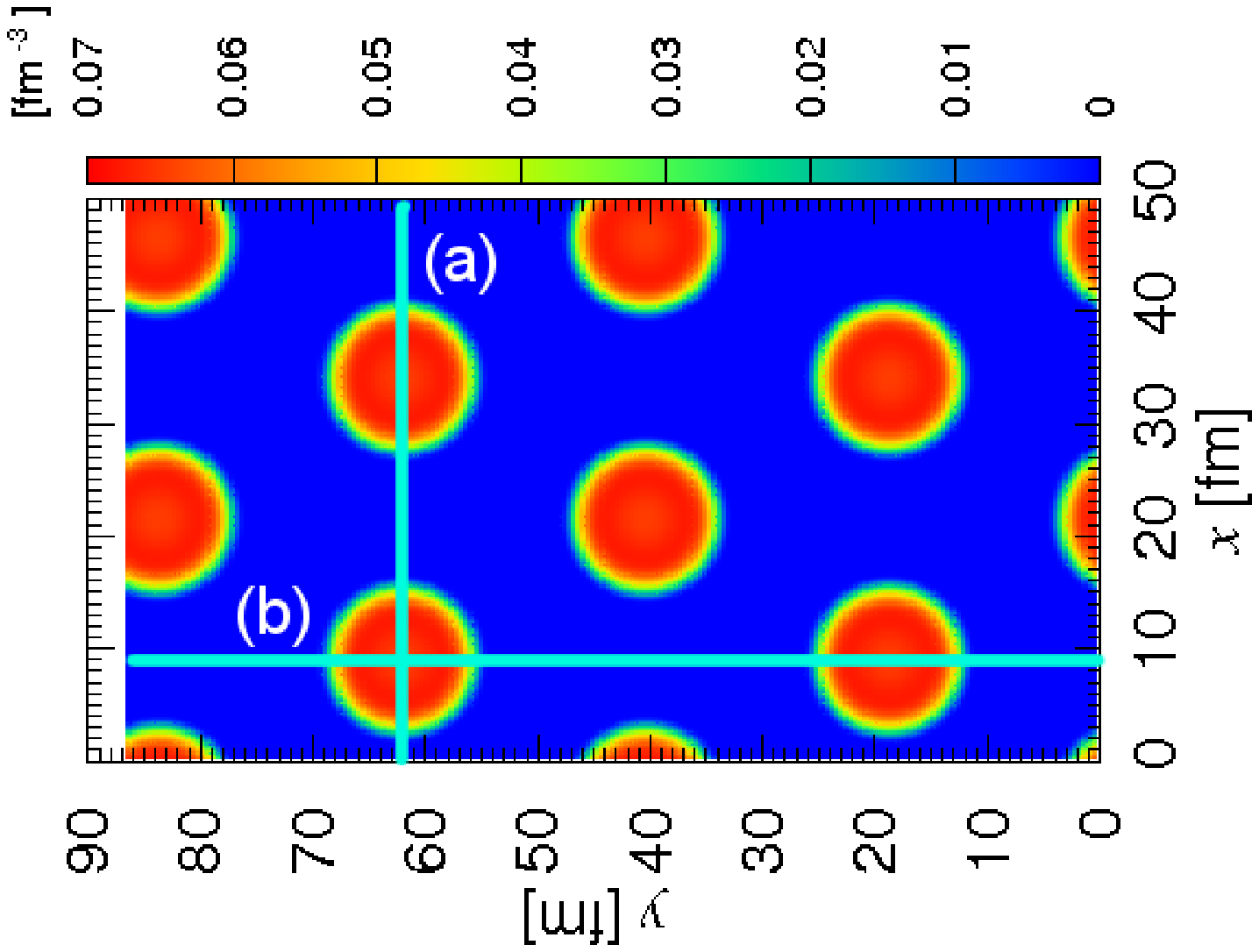}}
  \subfigure{\includegraphics[height=0.8\linewidth,angle=-90]{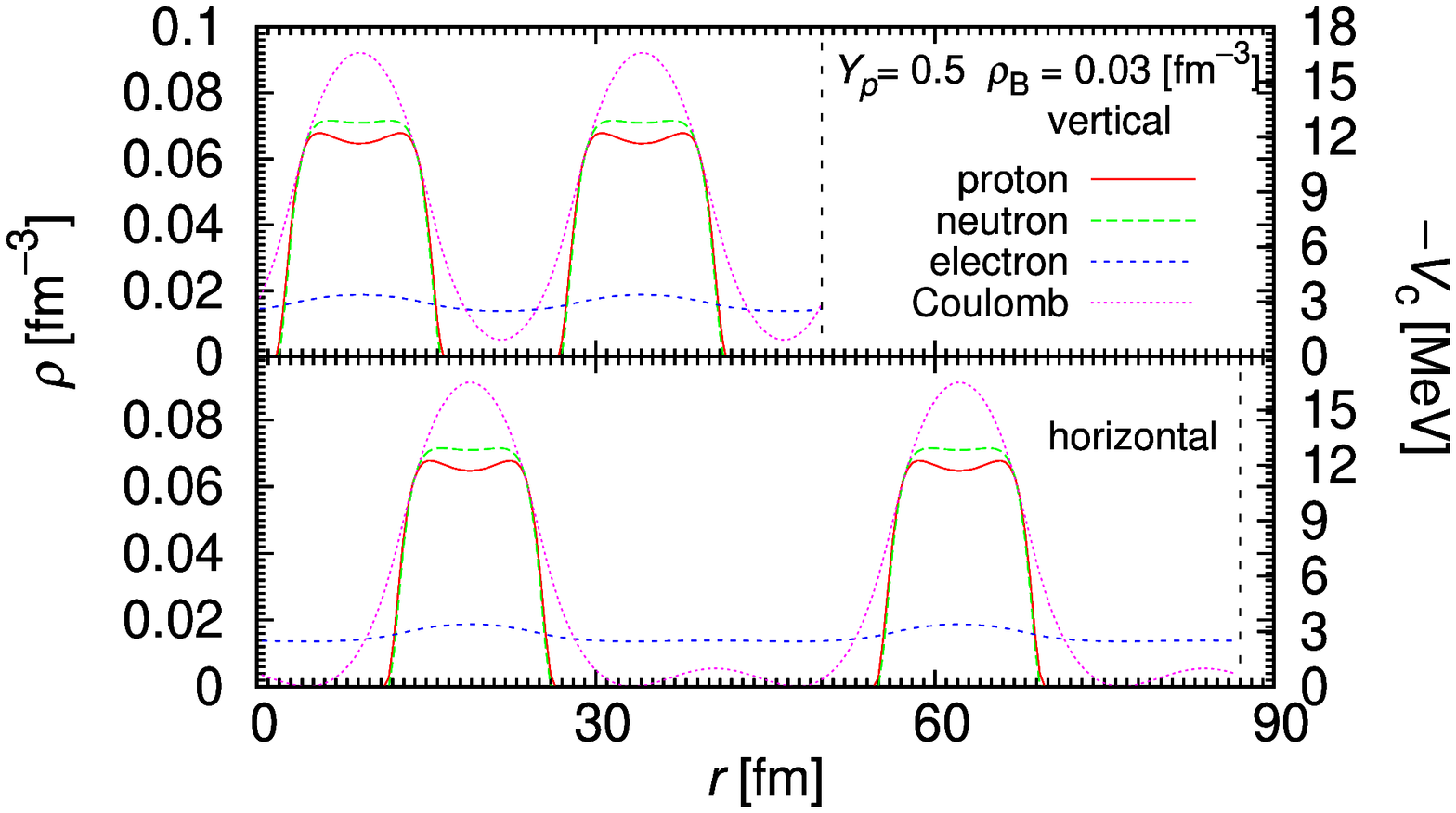}}
  \caption{(color online) Upper: Proton density distribution in the rod phase ($\rho_B=0.03$ fm$^{-3}]$ for $Y_p=$0.5) on the sliced plane.\\
  Lower: the density distribution along a line (a)(vertical) and (b)(horizontal) in upper figure.
  Red lines indicate proton, green neutron, blue electron and pink the Coulomb potential.}
  \label{dp_Yp=05}
 \end{center}
\end{figure}

Any exotic mixture does not appear as a ground state at any density. 
In a droplet, we have seen that the proton density is highest near the surface due to the Coulomb repulsion, 
while the neutron density distribution is flat inside the droplet. 
Note that baryon density outside the droplets is zero for $Y_p$=0.3 and 0.5. 
Electron density is spread over all space but slightly localized around the droplets 
which brings about the charge screening effect.

We can see the density distribution of fermions for $Y_p=0.5$ and $\rho_B=0.03$ fm$^{-3}$ in Fig.\ \ref{dp_Yp=05} and find 
the unique features of the three-dimensional calculation. 
Here, in the upper panel, we show the proton density distribution on a sliced plane and depict the two kinds of the density profiles of proton, 
neutron, electron and the Coulomb potential in the lower panel. 
One is along a vertical line (a) which passes through the rods, 
another is along a horizontal one (b). 
From this figure, the advantage of the three-dimensional calculation can be seen. 
The proton, neutron, and electron density distributions are almost the same for both the cases of (a) and (b). 
However, a slight difference appears in the Coulomb potential. 
We have set the maximum value of the Coulomb potential to be zero for convenience. 
In the case of the rod phase, that point corresponds to the centroid of the triangular lattice. 
These points are included not on the path (a) but on the path (b). 
Considering the importance of the distinct relation between the Coulomb and surface energies for the pasta structure, 
we should take into account this anisotropy in a proper way.

In Fig.\ \ref{EOS_fixed} we show the energy,  total pressure, and  baryon partial pressure as functions of density. 
Baryon partial pressure is given by subtracting the electron contribution from the total pressure. 
Note that the energy $E/A-m_N$  includes the kinetic energy of electrons, which makes the total pressure positive. 
This density dependence is qualitatively the same as the one with the WS approximation. 
The appearance of non-uniform structures will make nuclear matter more stable: 
the energy per baryon gets lower up to about 15 MeV/$A$ compared to uniform matter 
and the pressure per baryon higher up to about 0.5 MeV/$A$.

We have obtained almost the same EOS with that given by the WS approximation, in which the same RMF model is applied. 
However, one of the differences between our results and those with the WS approximation appears in the existing region of each pasta structure; 
the density region of the rod is wider and the tube narrower in our calculation. 
Since the energy differences between different structures are quite small,  
the crystalline configuration might affect the appearance of each pasta structure.

\begin{figure*}[htpb]
 \begin{minipage}{0.8\textwidth}
  \subfigure{\includegraphics[height=0.32\linewidth,angle=-90]{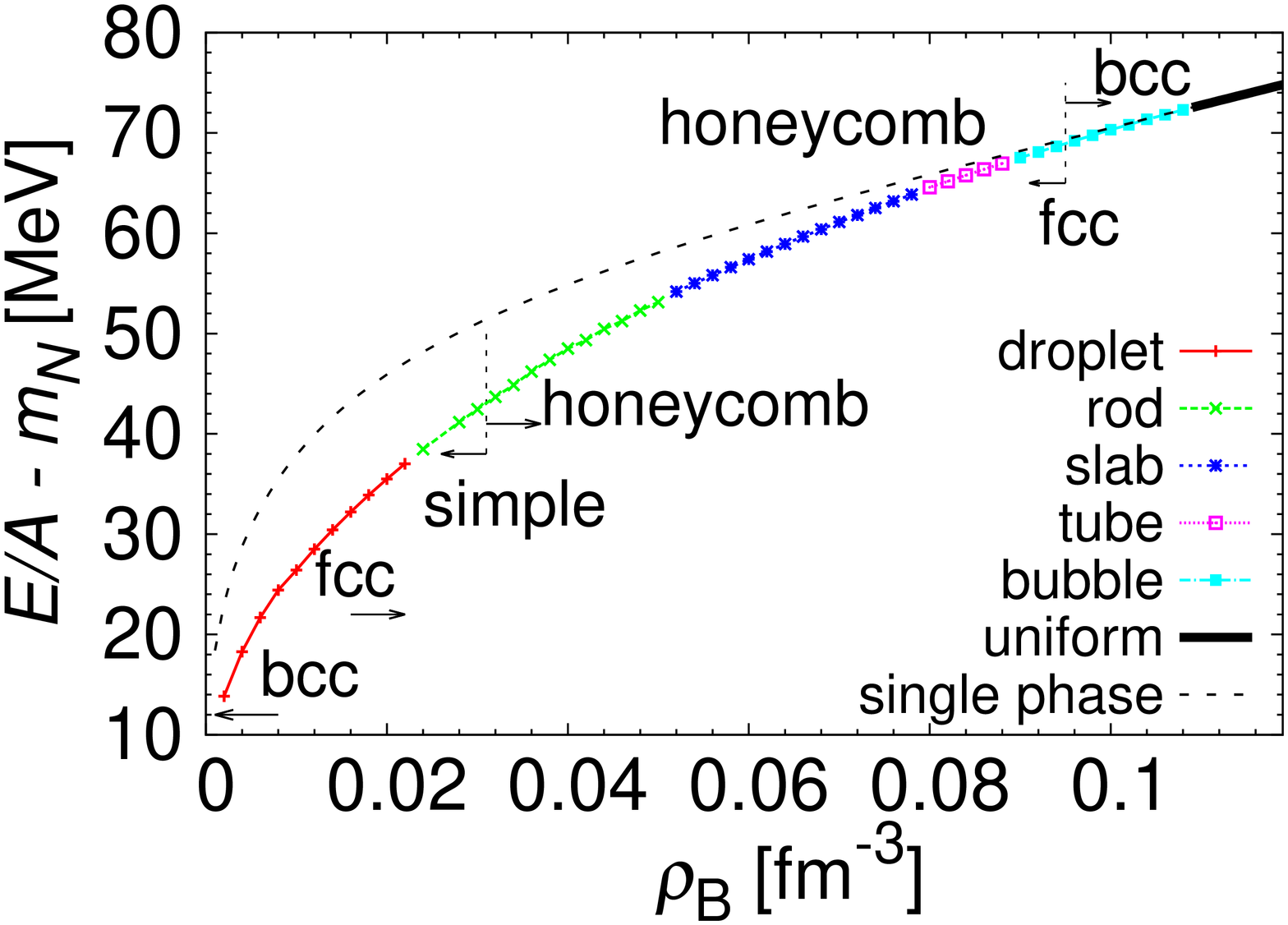}}
  \subfigure{\includegraphics[height=0.32\linewidth,angle=-90]{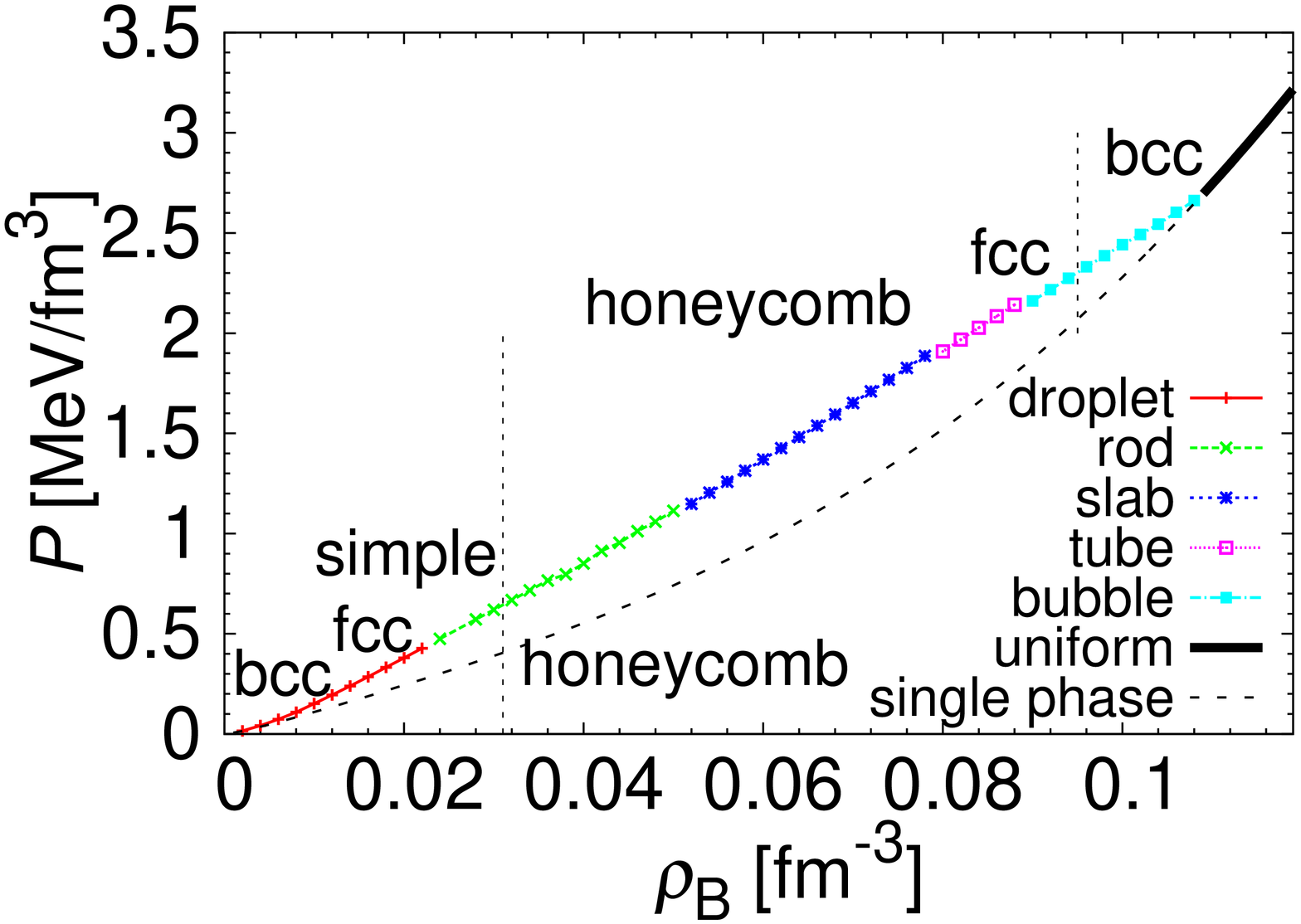}}
  \subfigure{\includegraphics[height=0.32\linewidth,angle=-90]{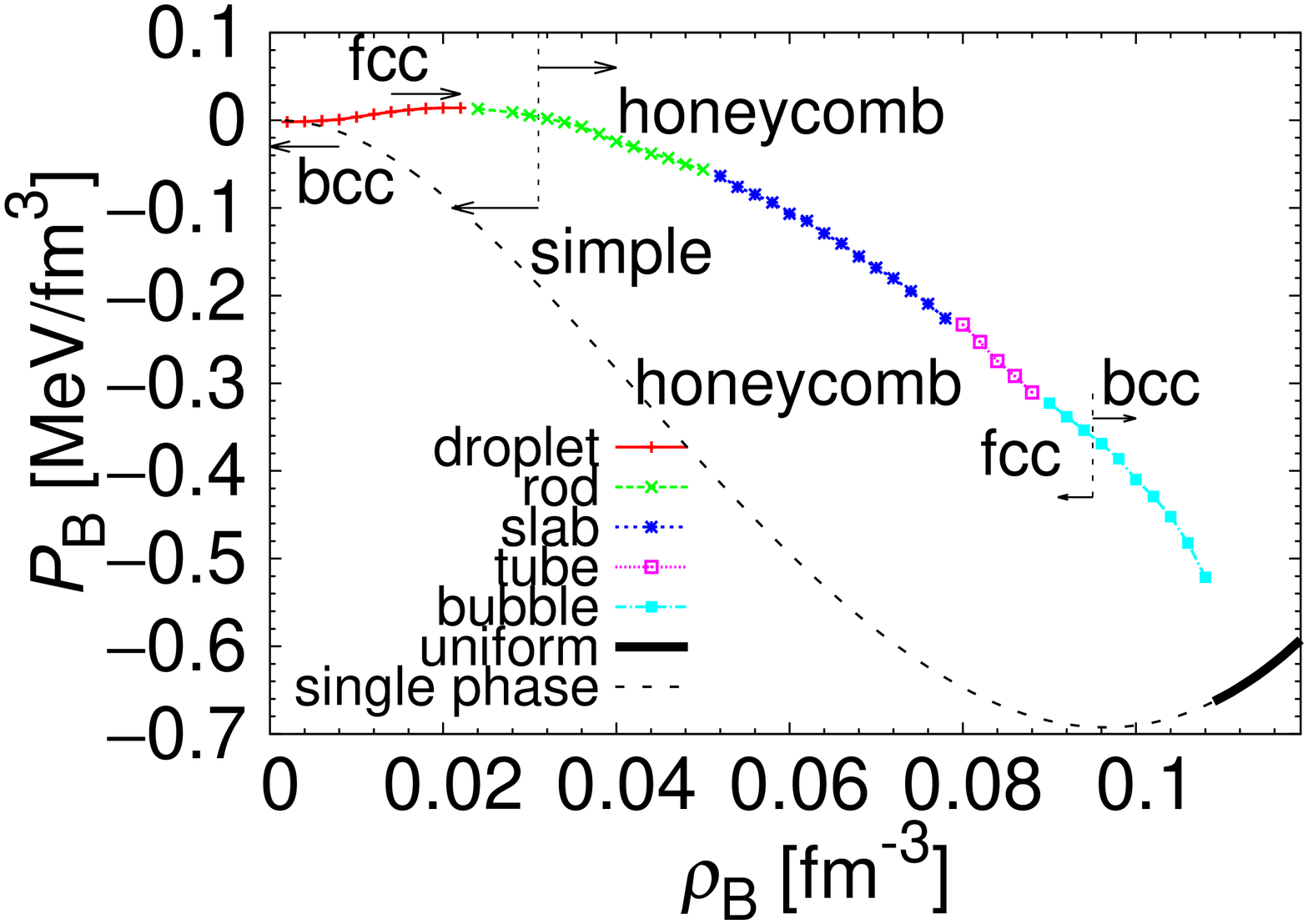}}
\end{minipage}
 \hspace{-5.0mm}
 \begin{minipage}{0.8\textwidth}
  \subfigure{\includegraphics[height=0.32\linewidth,angle=-90]{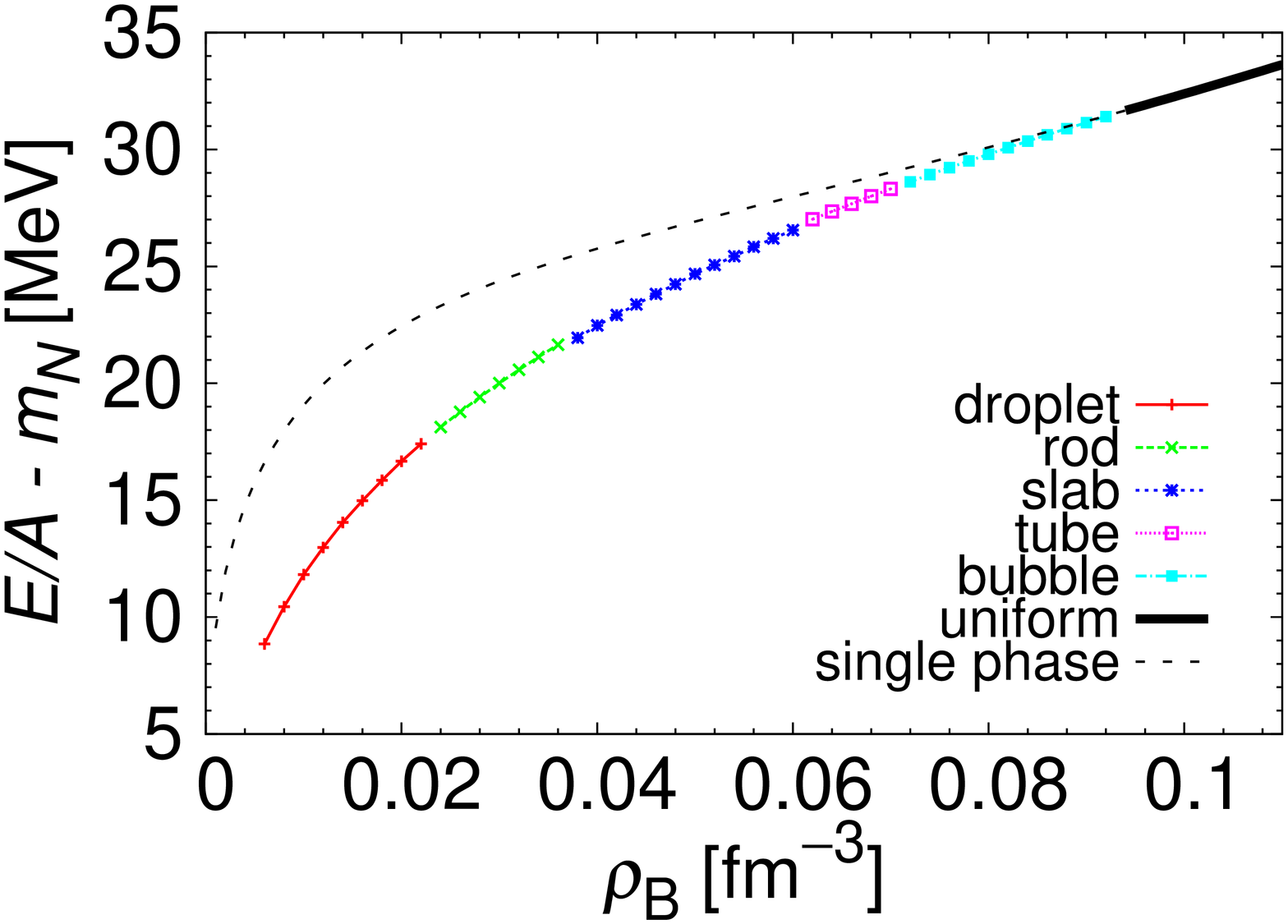}}
  \subfigure{\includegraphics[height=0.32\linewidth,angle=-90]{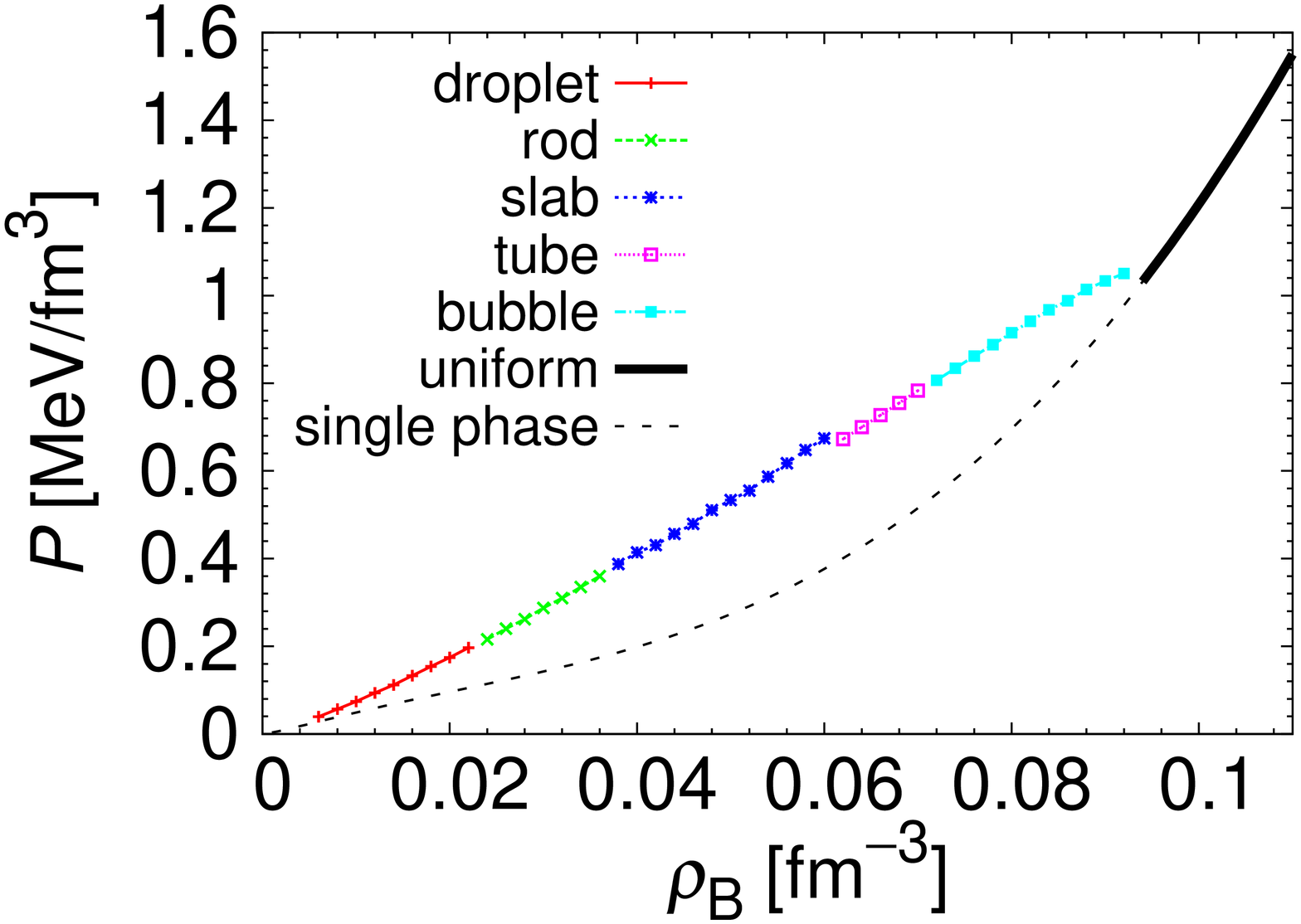}}
  \subfigure{\includegraphics[height=0.32\linewidth,angle=-90]{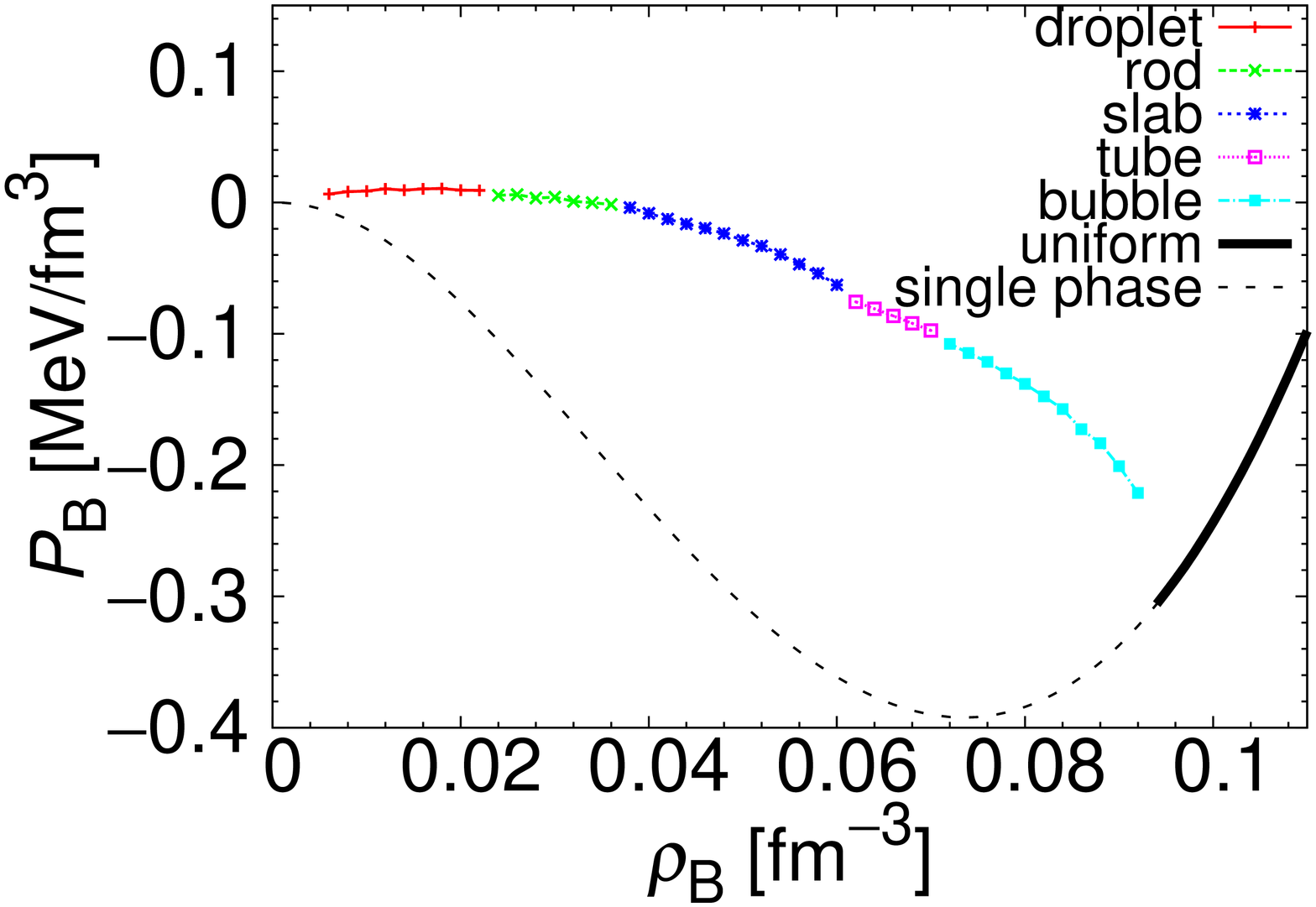}}
 \end{minipage}
 \begin{minipage}{0.8\textwidth}
  \subfigure{\includegraphics[height=0.32\linewidth,angle=-90]{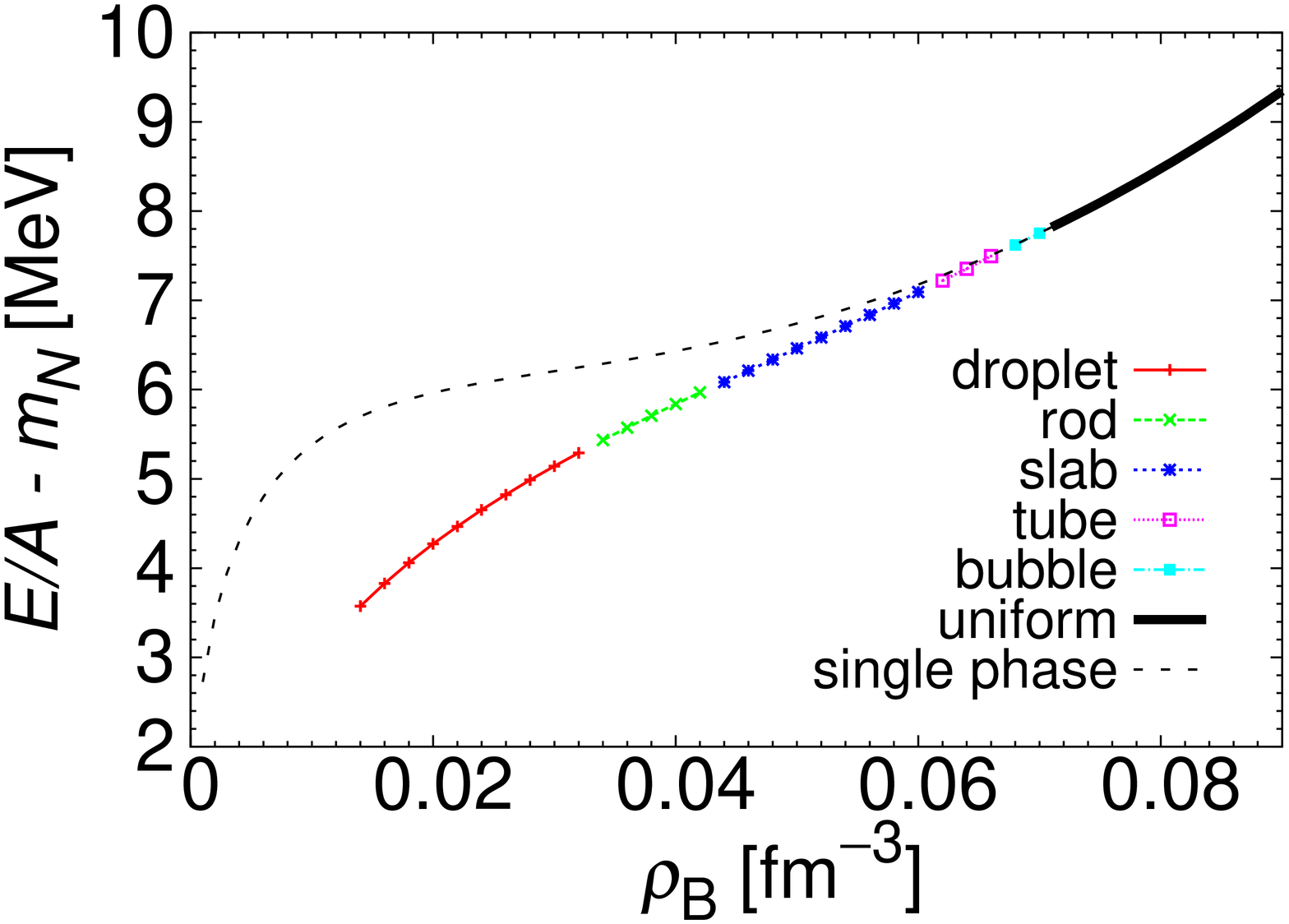}}
  \subfigure{\includegraphics[height=0.32\linewidth,angle=-90]{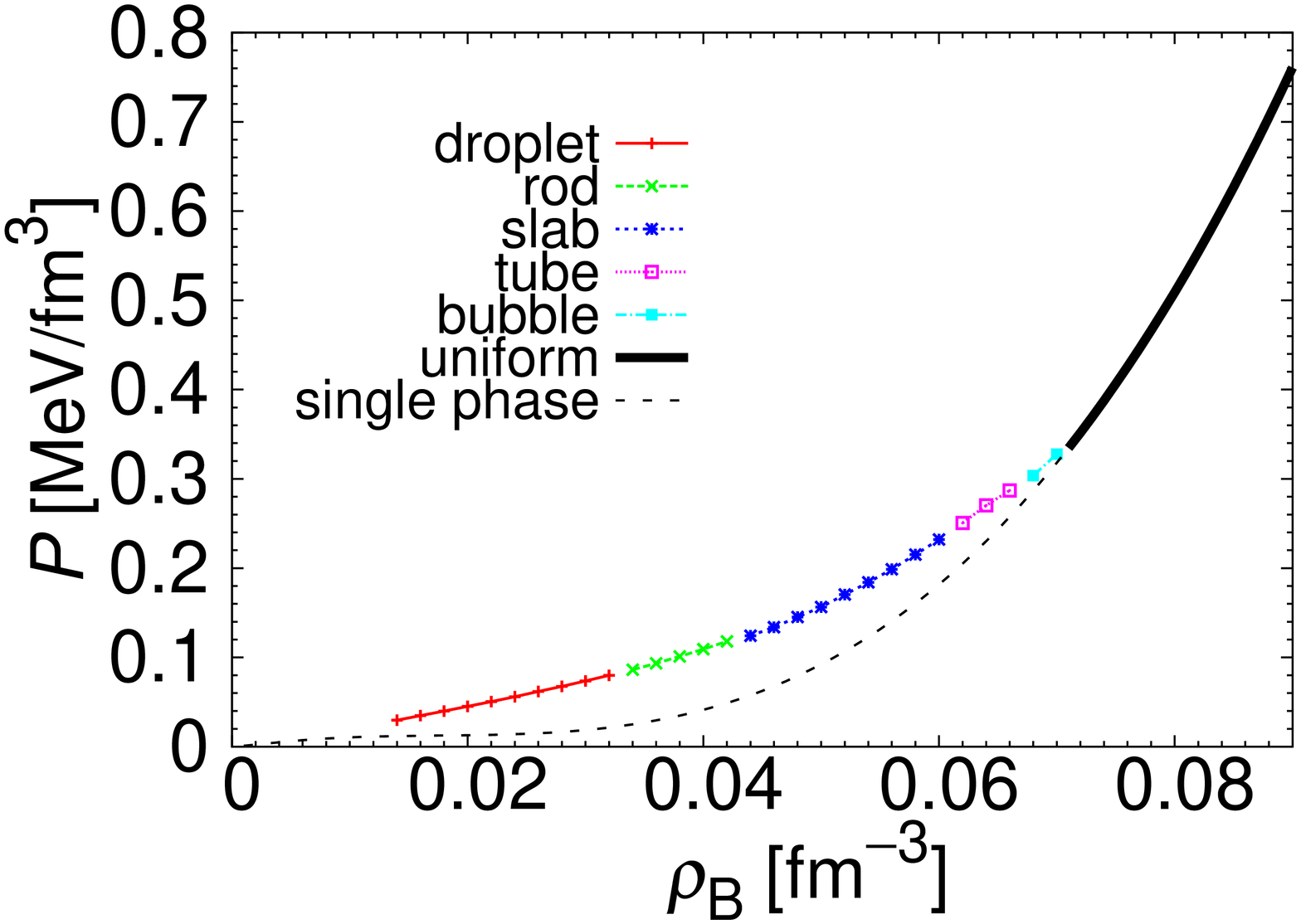}}
  \subfigure{\includegraphics[height=0.32\linewidth,angle=-90]{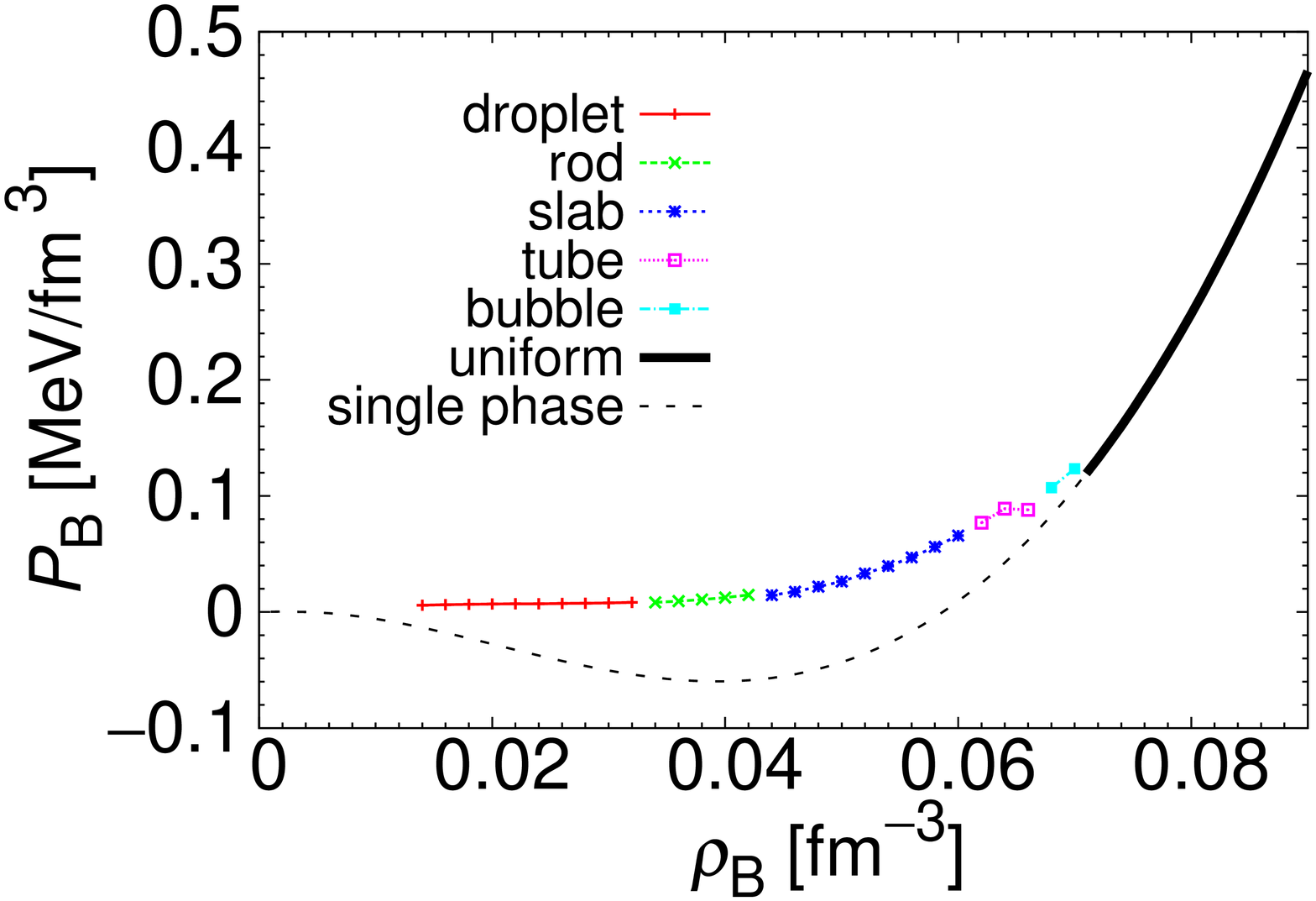}}
 \end{minipage}
  \caption{(color online) From the left, energy, pressure, and baryon partial pressure 
   of $Y_p$=0.5, 0.3, 0.1 in this order from the  upper panel.
   Red lines indicate droplet, green rod, blue slab, magenta tube, cyan bubble, and black uniform, respectively.}
 \label{EOS_fixed}
 \end{figure*}

\begin{figure}[htbp]
 \begin{center}
  \includegraphics[width=0.5\linewidth,angle=-90]{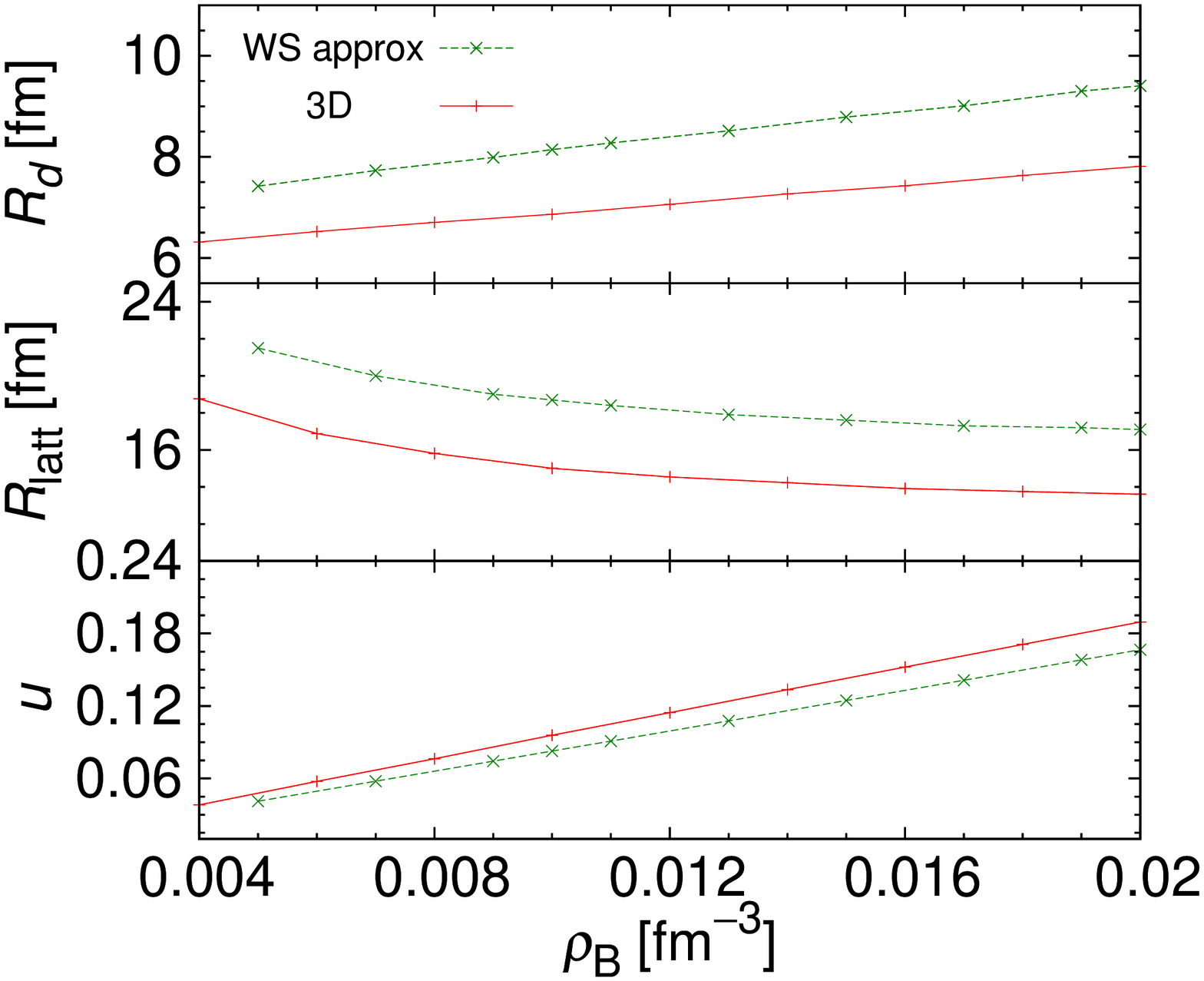}
 \end{center}
 \caption{(color online) Density dependence of the radius, the lattice constant, and the volume fraction.
  Red lines are our results and green lines are those with the WS approximation.}
 \label{size}
\end{figure}

Figure \ref{size} shows the radius of droplets $R_d$, the lattice constant $R_{\rm latt}$, and the volume fraction $u$ in the droplet phase. 
Here, $R_d$ and $R_{\rm latt}$ are defined as follows, 
\begin{eqnarray}
\frac{V}{N_d} &=& \frac{4{\pi}}{3}R_{\rm latt}^3, \label{Rcell} \\
 R_d &=& R_{\rm latt}\left(\frac{\langle{\rho}_p\rangle^2}
{\langle{\rho}^2_p\rangle}\right)^{1/3}, \label{Rd}
\end{eqnarray}
and $u=(R_d/R_{\rm latt})^3$, 
where $V$ denotes the cell volume, 
$N_d$ the number of droplets in the cell, and the bracket $\langle ... \rangle$ means the average over the cell volume. 
Comparing the present results with those using the WS approximation, the volume fraction exhibits the same behavior, 
but the lattice constant and the radius of the droplet the different behavior. 
This might be caused by some differences of the treatment of the Coulomb interaction between the present calculation and the one 
using the WS approximation.

Although any unusual structure does not appear in the ground state, 
one new feature emerges in the crystalline configuration of droplets. 
It has been considered within CLDM that a body-centered cubic (bcc) lattice is energetically favored 
than the face-centered cubic (fcc) lattice by the Coulomb energy \cite{oyamatsu}.
However, in our calculation, the fcc lattice emerges as a ground state at some density region.
Crystalline configurations of the bcc and fcc lattices give rise to a subtle difference of the Coulomb energy, 
which amounts to about $0.2$--$0.8$ MeV. 
The ratio of the Coulomb energy to the total energy difference is about $20\%$. 
In other words, the major contribution to the energy  comes from the mean-fields and the kinetic energies of nucleons,
i.e., the bulk energy of nuclear matter. 
Therefore, the ansatz used in the previous studies, e.g., CLDM, should be dangerous since small differences in size and composition of droplets give a significant change in the total energy of the system.
\begin{table}[htbp]
\begin{ruledtabular}
 \begin{tabular}{cccccc}
  $\rho_B$ [ fm$^{-3}$] &   0.012  &  0.014  &  0.016  &  0.018  &  0.020 \\
  \hline
  $R_d(\rm fcc)$ [fm]    &  6.86   &  7.04  &  7.23  &  7.61  &  7.79 \\
  $R_d(\rm bcc)$ [fm]    &  6.99   &  7.18  &  7.36  &  7.75  &  7.92 \\
 \end{tabular}
\end{ruledtabular}
 \caption{Density dependence of the droplet radii for the fcc and bcc
 lattices.}
 \label{dif_f_b}
\end{table}
Table \ref{dif_f_b} shows the density dependence of radii of the droplets in the cases of the fcc and bcc lattices. 
The radius of the droplet is different even if their baryon densities are the same. 
In Refs.\ \cite{oyamatsu}, it has been reported that the bcc lattice is realized 
in the ground state, 
by comparing the bcc and fcc crystals of 
droplets of the same size, while they have different droplet sizes in our case. 

In the QMD calculations \cite{qmd1} that precede the present calculation 
without assuming geometrical structures, 
droplets form the bcc lattice. 
This difference might come from the treatment of electrons or the charge screening effect: 
uniform electron distribution has been assumed in the QMD calculation. 
To see the effects of the electron distribution on the crystalline configuration,
let us compare two cases within our framework: one is the full calculation and the other assumes uniformly distributed electrons. 
In the latter calculation, the Coulomb potential $V_{\rm Coul}$ in Eq.(\ref{eq:rhoe}) is 
replaced by a constant $V_0 = 0$ and ${\rho_e} =    {(\mu_e - V_0)^3} / {3\pi^2}$.
In the full calculation using (\ref{eq:rhoe}), the value of $V_0$ is arbitrary due to the gauge invariance, 
and one can take $V_0$ for the sake of convenience, e.g.\ , as $V_0$ = 0. 
However, in the case of uniformly distributed electrons, the gauge invariance is partially violated, 
since we replace $V_{\rm Coul}$ to $V_0$ in the equation for the electron chemical potential 
but retain $V_{\rm Coul}$ in the equation for the proton chemical potential and thus in the expression for the proton number density. 
In this case, the droplets have a smaller size compared with that in the full calculation in all the density region of the droplet phase. 
This means that the Coulomb repulsion  among  protons is slightly weaker   in the full calculation due to the screening by electrons. 
One can see this difference in Table \ref{Coul_dif}. 
\begin{table}[htbp]
\begin{center}
\begin{ruledtabular}
 \begin{tabular}{cccccc}
  $\rho_B$ [fm$^{-3}$] &  0.004  &  0.01 & 0.016 & 0.022 \\
  \hline
  $R_d$ (full calc) [fm] & 6.09 & 6.67  &  7.23  &  7.79 \\
  $R_d$ (unif elec) [fm] & 5.91 & 6.40  &  7.01  &  7.64
 \end{tabular}
\end{ruledtabular}
 \caption{The radii of droplets by the full calculation and with
          uniformly distributed electrons.}
 \label{Coul_dif}
\end{center}
\end{table}
In our previous study using the WS approximation \cite{maruyama}, a similar discussion was made. 
We argued that the charge screening effect is not so remarkable due to the large     Debye screening length. 
However,  the crystalline configuration does not change and the fcc lattice is more favored. 
Thus, it is confirmed that the charge screening by electrons does not very much affect the crystalline configuration. 
This difference of the crystalline configuration between the QMD calculation and the 
present calculation remains to be elucidated as a future problem.

We obtain the typical pasta structure as a ground state for any proton number-fraction above 0.1. 
Also in the cases of $Y_p$=0.3 and 0.1, the fcc lattice of droplets is energetically more favorable than the bcc one. 
In Fig.\ \ref{dp_Yp=01} we depict the density profiles of proton, neutron, electron and the Coulomb potential for $Y_p= 0.3$ and 0.1 
with baryon number density $0.03$ fm$^{-3}$ 
along a line which passes 
through the rods in the same way as in Fig.\ \ref{dp_Yp=05}. 
While for $Y_p=$0.3 there appear vacant regions of neutron, the neutron density is finite at any point for $Y_p=$0.1: 
the space is filled with dripped neutrons. 
\begin{figure}[htbp]
 \begin{center}
  \subfigure{\includegraphics[height=0.9\linewidth,angle=-90]{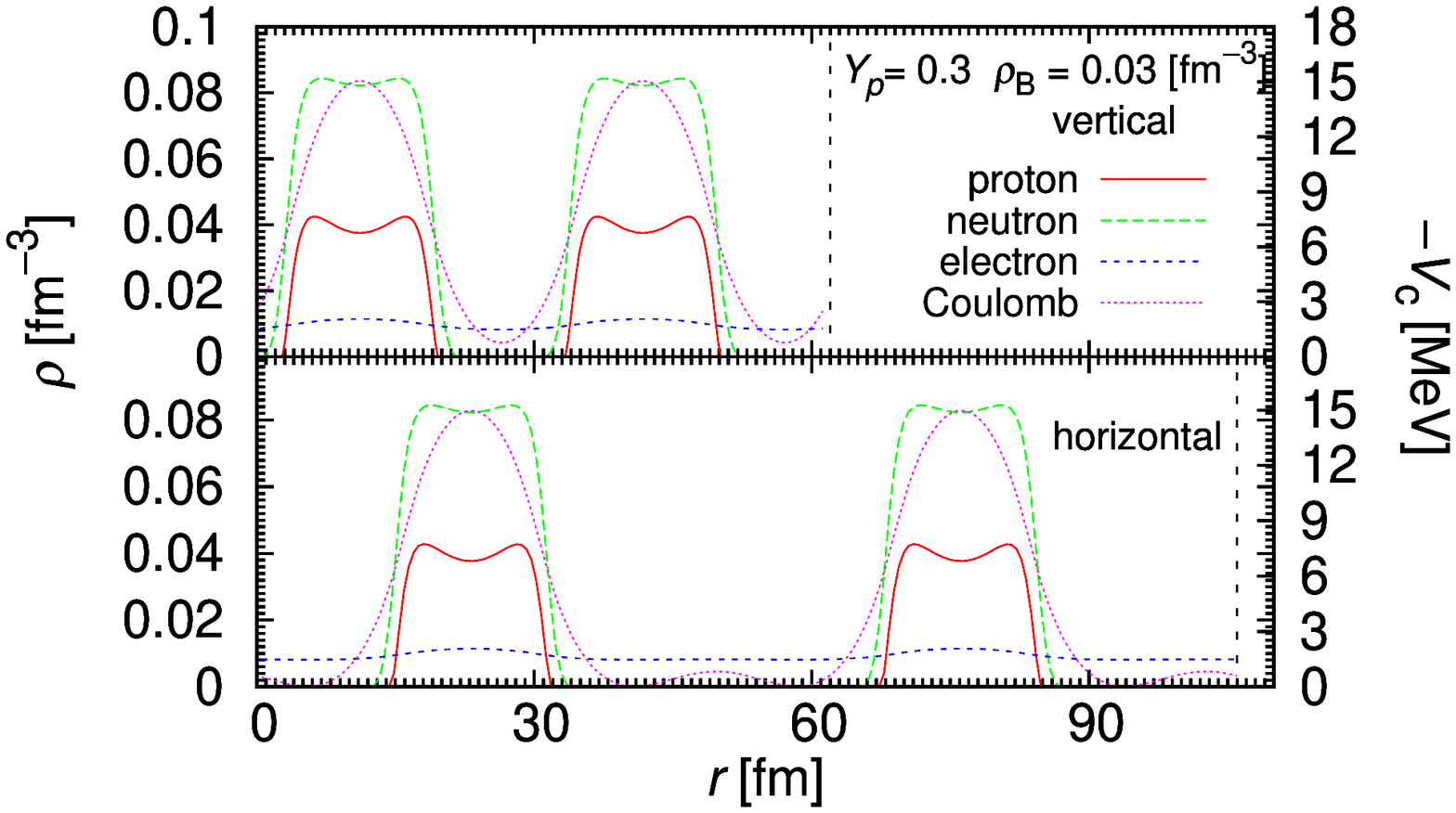}}
  \subfigure{\includegraphics[height=0.9\linewidth,angle=-90]{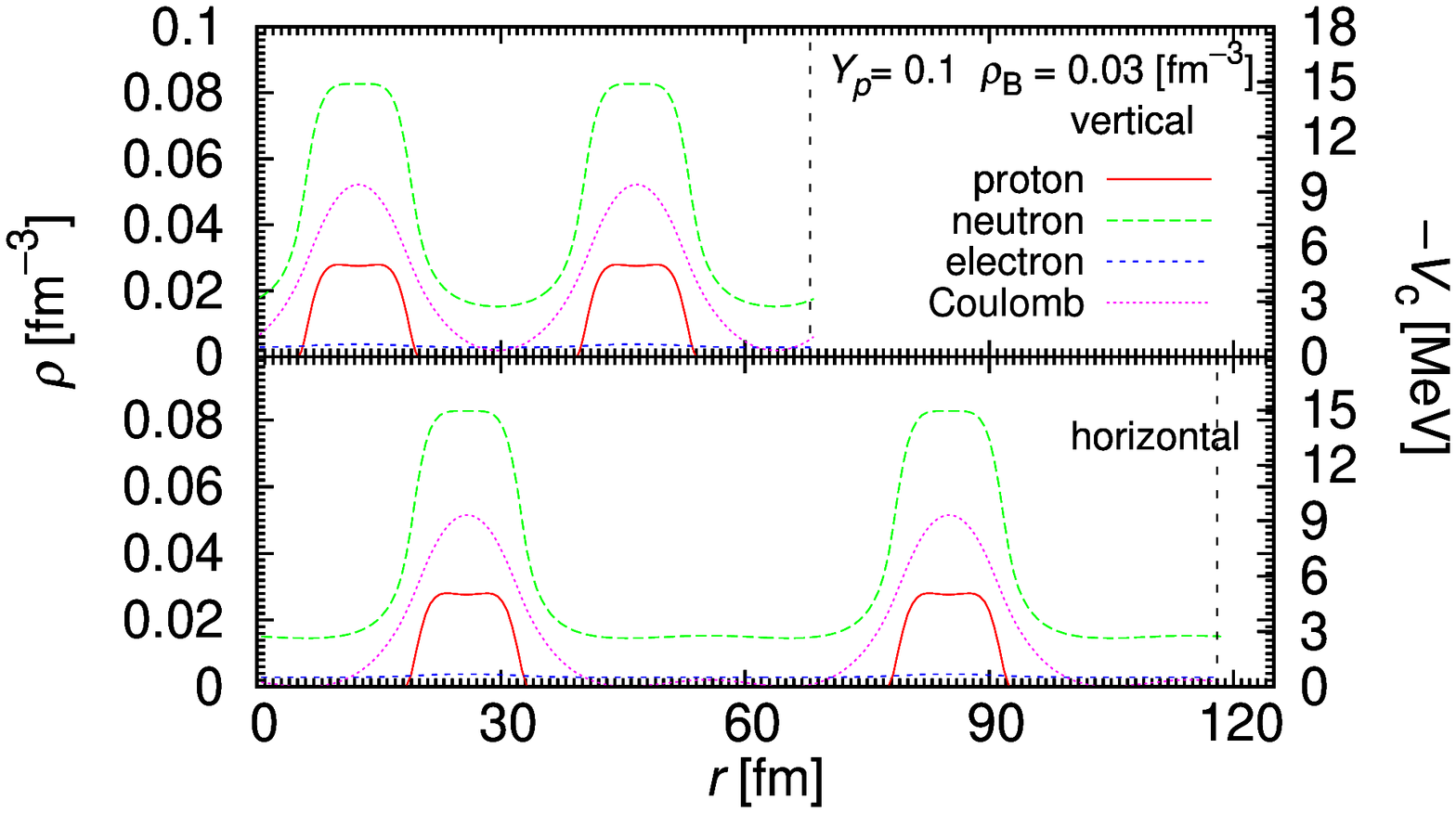}}
  \caption{
  (color online) Density distribution of proton, neutron, electron and the Coulomb potential of rod in $\rho_B=0.03$ fm$^{-3}$
  (upper: $Y_p=0.3$, lower: $Y_p=0.1$)
  Red lines indicate proton, green neutron, blue electron, and pink the Coulomb potential.}
  \label{dp_Yp=01}
 \end{center}
\end{figure}
Even    in the case of $Y_p=0.1$, 
we can see that the proton density is highest around the surface due to the Coulomb repulsion.

\subsection{Catalyzed matter}
Cold catalyzed matter requires beta equilibrium instead of the fixed proton number-fraction: $\mu_n=\mu_p+\mu_e$.
Shown in Fig.\ \ref{pasta_beta} are the proton density distributions of the ground states in cold catalyzed nuclear matter. 
We have obtained the body-centered cubic (bcc) lattice of droplets, face-centered cubic (fcc) lattice of droplets, and 
honeycomb lattice of rods, depending on density. 
In Fig.\ref{distribution}, the density profiles of fermions in the bcc and fcc 
lattices of droplets and the Coulomb potential are depicted 
along a line which
passes through the droplets for $\rho_B\approx 0.01$ fm$^{-3}$ and $\rho_B\approx 0.03$ fm$^{-3}$. 
The effect of the Coulomb repulsion can be seen where the proton density is highest near the surface, 
while the neutron density distribution is flat in the droplet. 
We have observed only those three structures (a), (b), (c) in Fig.\ \ref{pasta_beta}. 
This result is consistent with the previous study about the relation between the density region of the  
pasta structure and the slope parameter $L$ in Eq.~(10) \cite{iida}. 
The larger $L$ is given, the narrower pasta region is. 
In our calculation, $L$ is about $80$ MeV, 
which is close to the critical value $\approx  90$ MeV in the CLDM, where the pasta structures do not appear. 
In the previous calculation using the WS approximation, only the droplet structure appeared as a ground state \cite{maruyama}. 
However, in our three-dimensional calculation using the same RMF framework,  the rod structure also appears.

\begin{figure}[htpb]
 \begin{minipage}{0.5\textwidth}
    \includegraphics[width=\textwidth]{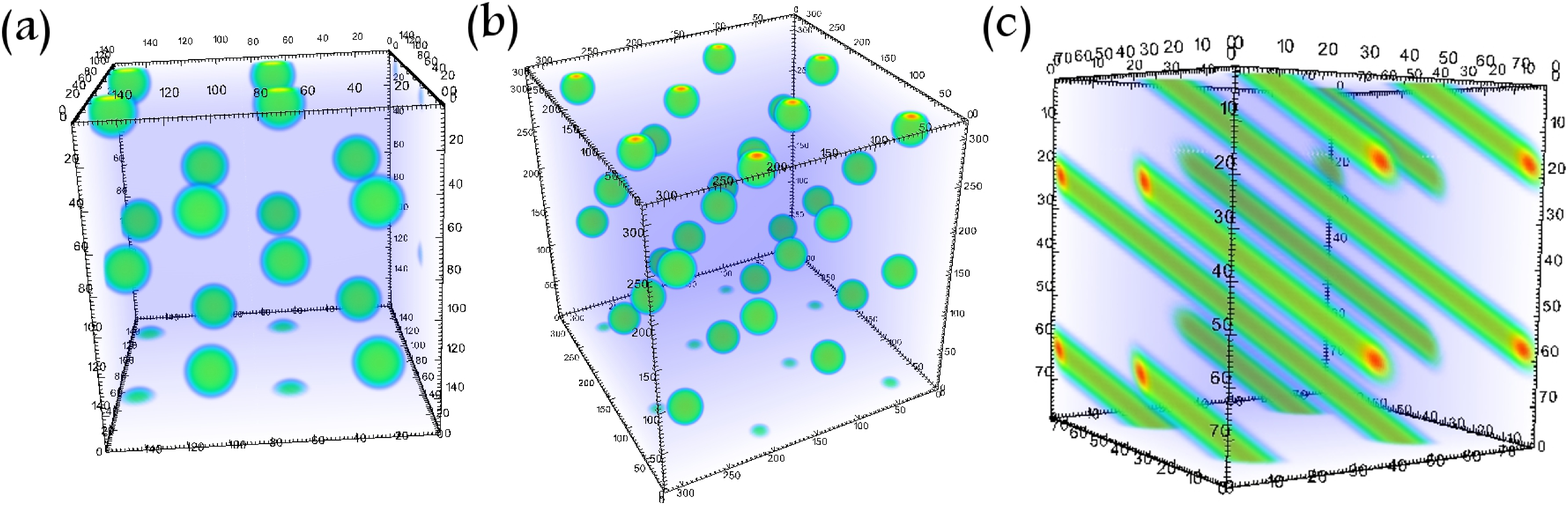} 
  \caption{
  (color online) Proton density distributions are shown; droplet (bcc)
  at $\rho_B=0.01$ fm$^{-3}$, droplet (fcc) at $\rho_B=0.03$ fm$^{-3}$, 
  rod (honeycomb) at $\rho_B=0.056$ fm$^{-3}$.} \label{pasta_beta}
 \end{minipage}
 \hspace{10mm}
 \begin{minipage}{0.49\textwidth}
  \includegraphics[width=\textwidth,angle=-90]{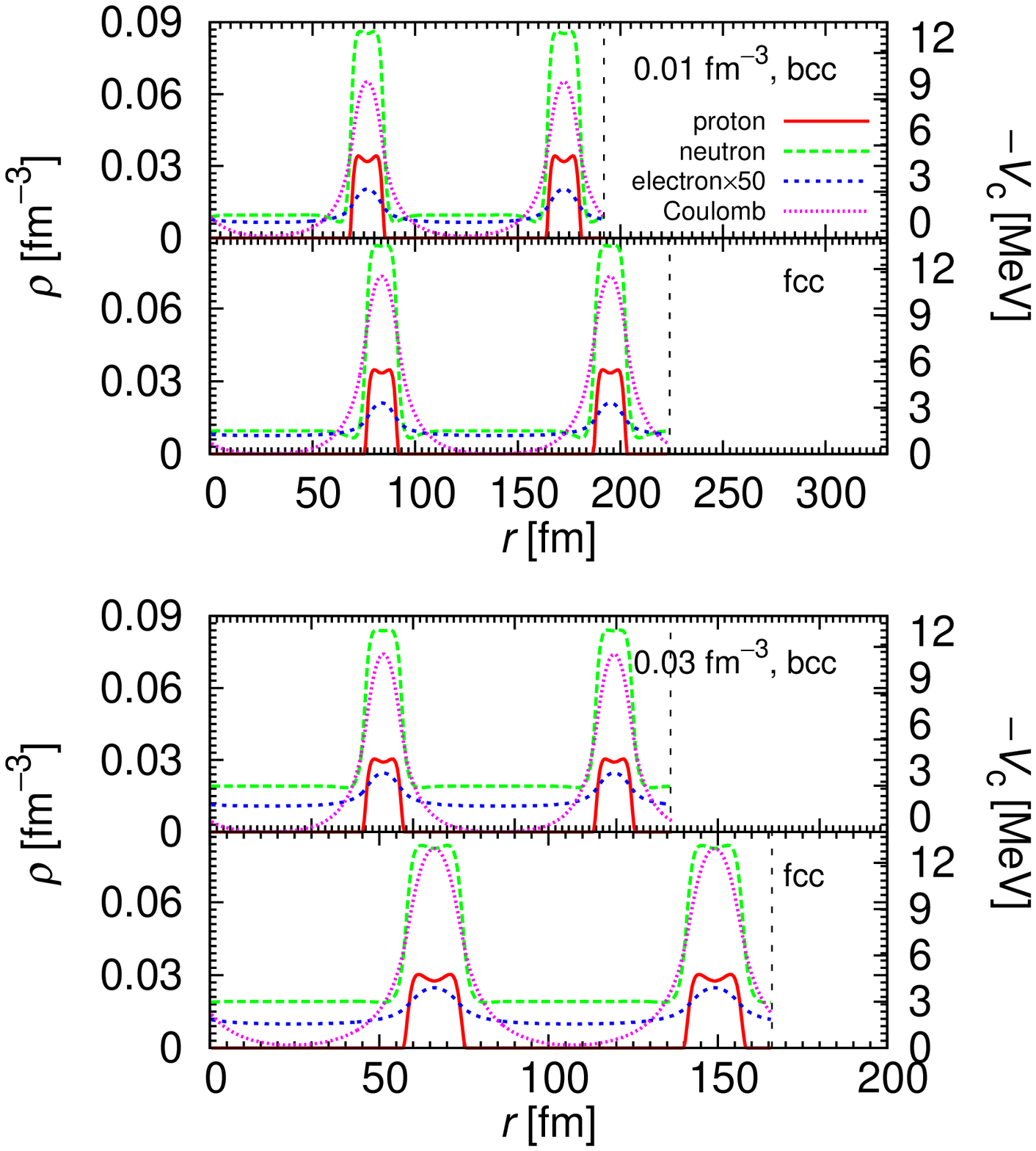}
  \caption{
  (color online) Density distributions of proton, neutron, and electron and the Coulomb
  potential in the fcc and bcc lattices. The upper panel shows the case of $\rho_B=0.01$
  fm$^{-3}$ and $\rho_B=0.03$ fm$^{-3}$ in the lower panel. Red lines
  indicate the proton, green neutron, blue electron (50 times) density
  distributions, pink the Coulomb potential, and black dashed lines are the lattice constant.} 
  \label{distribution}
 \end{minipage}
 \end{figure}

\begin{figure*}[htpb]
 \begin{center}
  \subfigure{\includegraphics[height=0.3\linewidth,angle=-90]{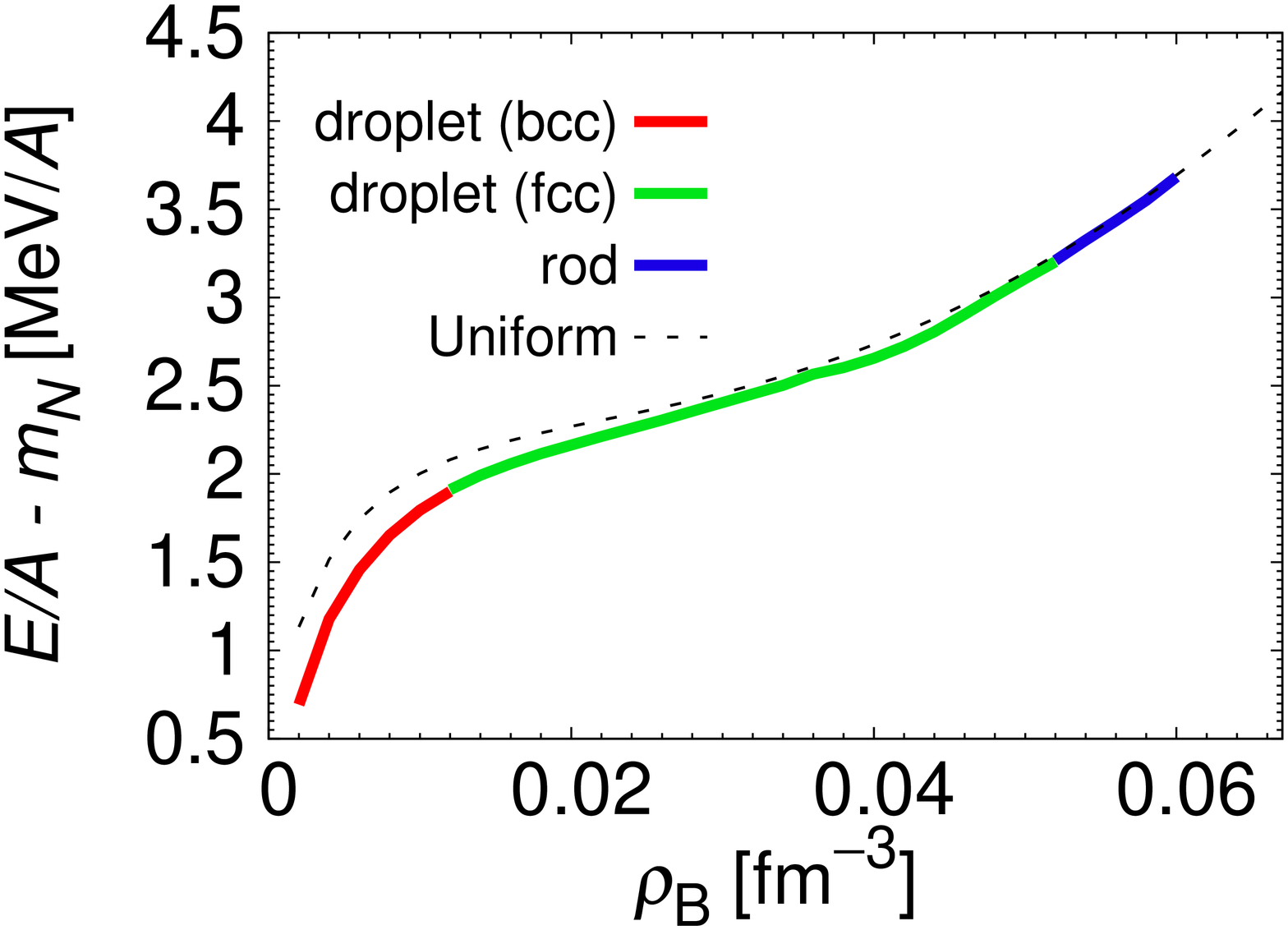}}
  \subfigure{\includegraphics[height=0.3\linewidth,angle=-90]{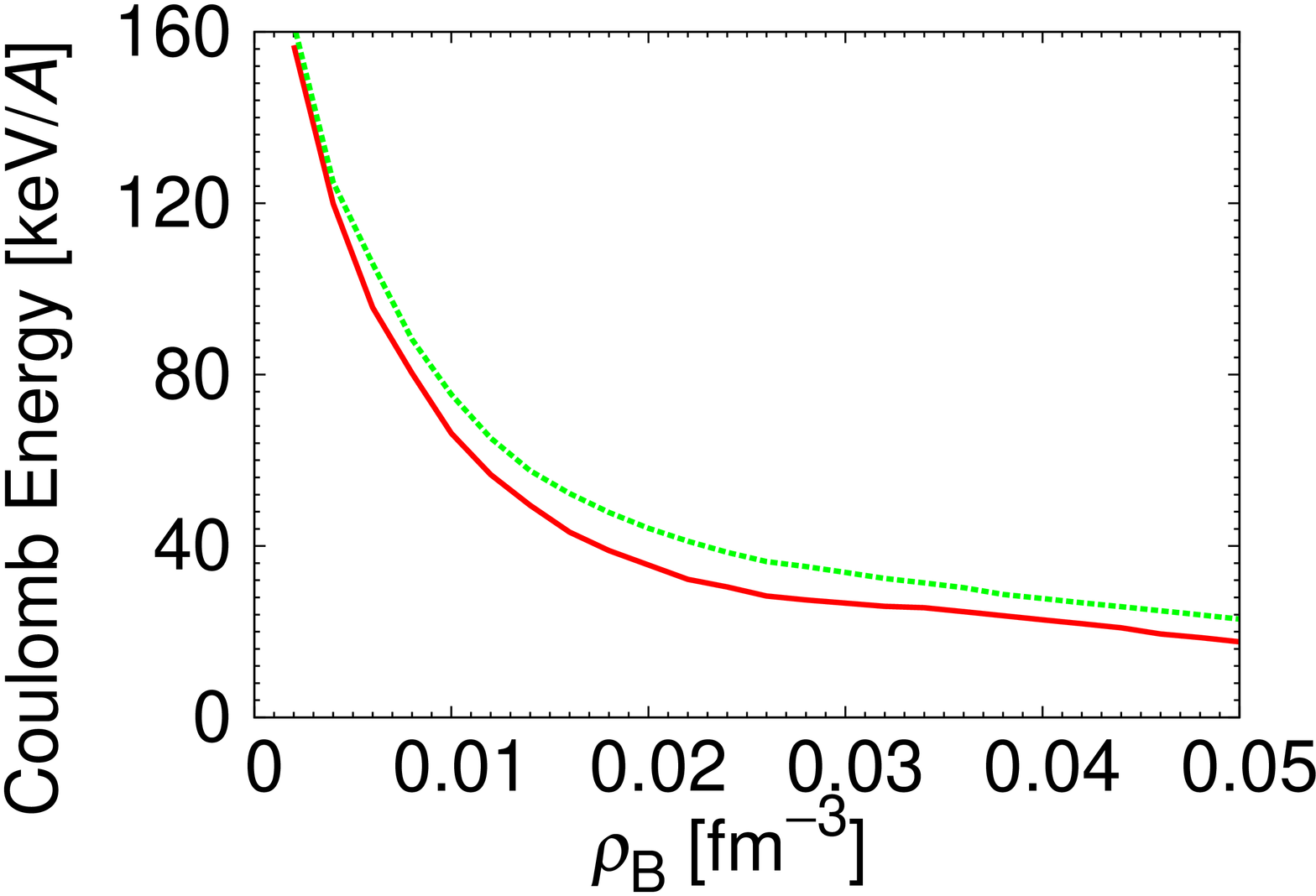}}
  \subfigure{\includegraphics[height=0.3\linewidth,angle=-90]{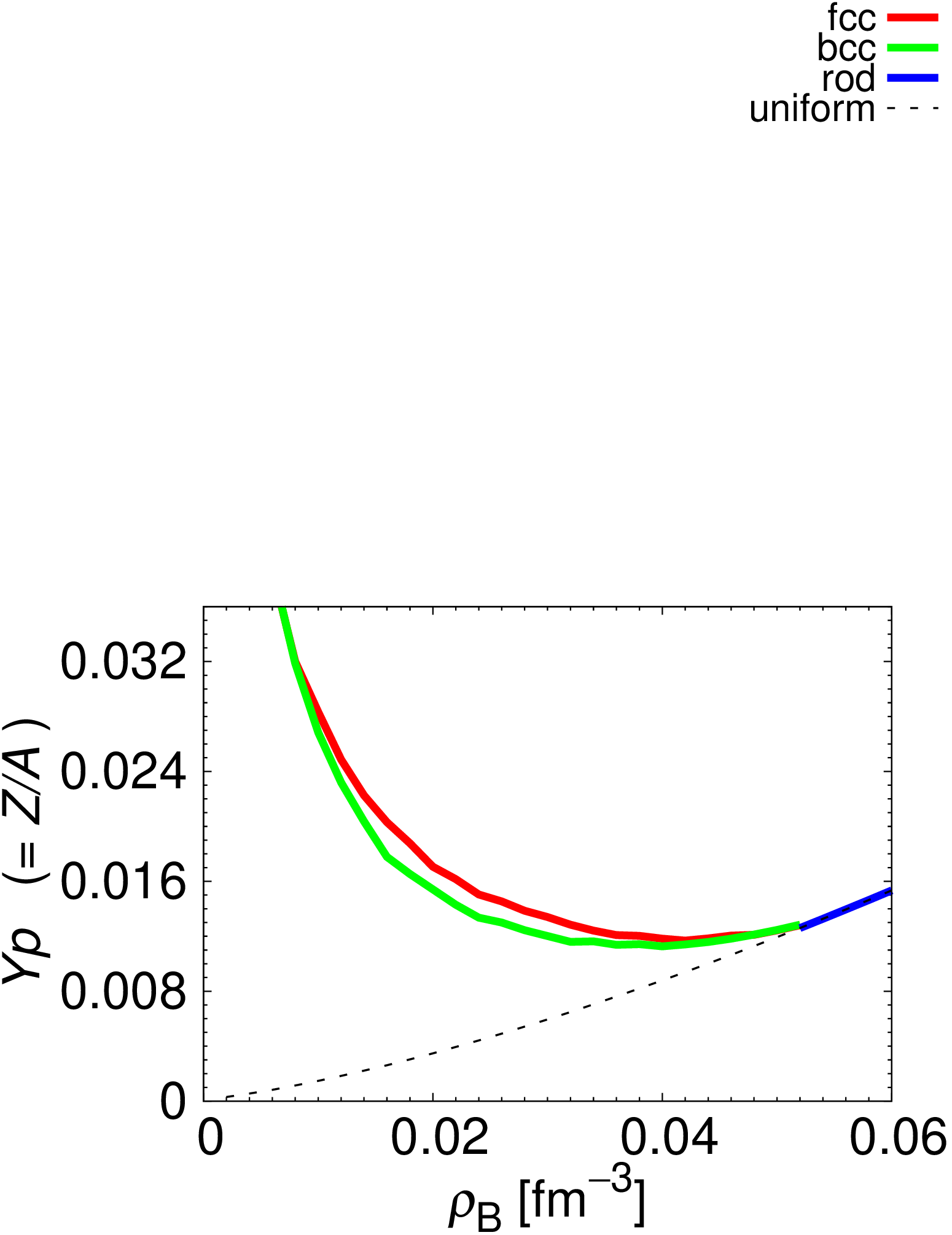}}
  \caption{
  (color online) From the left, the total energy, the Coulomb energy, and the 
  proton number-fraction.  Red lines indicate droplet (fcc),  green droplet (bcc), blue rod, respectively.
  Dotted lines are for the case of a single phase.}
  \label{EOS}
  \end{center}
\end{figure*}

We show the density dependence of the total energy, the Coulomb energy and the proton number-fraction in Fig.\ \ref{EOS}. 
To see the difference between droplets in the bcc and fcc lattices, we plot the density dependence of 
the size of droplet, lattice constant, volume fraction, proton number of each droplet and proton number-fraction in Fig.\ \ref{zyp}. 
Here, the meaning of  $R_d$ and $R_{\rm latt}$ are the same as in Eqs.\ (\ref{Rcell}) and (\ref{Rd}).

The density dependence of the ground state energy is shown in the left panel of Fig.\ \ref{EOS}. 
We can see that the ground state configuration changes depending on density. 
In the lower density region, the bcc lattice of droplets appear. 
Around $\rho_B\approx 0.01$ fm$^{-3}$, lattice structure changes from the bcc to fcc lattice. 
Remarkable change occurs around $\rho_B\approx 0.052$ fm$^{-3}$: 
from the fcc lattice of droplets to the honeycomb lattice of rods. 
It is hard to see the total energy difference between the bcc and fcc lattices of droplets. 
That of the proton number-fraction at $\rho_B<0.01$ fm$^{-3}$ is also hard to be distinguished in the right panel of Fig.\ \ref{EOS}. 
However, there are significant differences in the proton number-fraction at $\rho_B > 0.01$ fm$^{-3}$ and the Coulomb energy as shown 
in the middle panel of Fig.\ref{EOS}.

\begin{figure}[htbp]
 \begin{center}
  \subfigure{\includegraphics[height=0.49\linewidth,angle=-90]{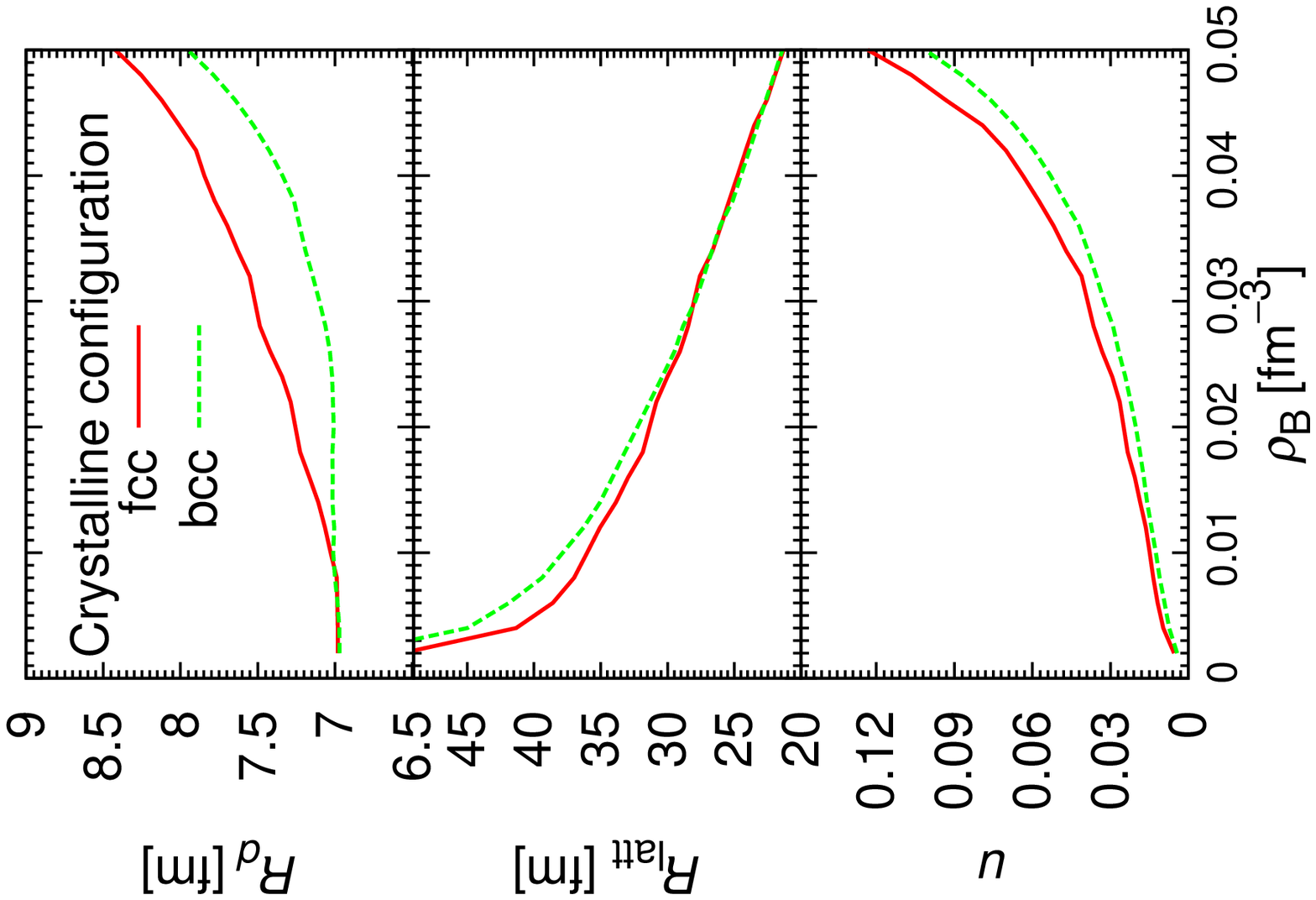}}
  \subfigure{\includegraphics[height=0.49\linewidth,angle=-90]{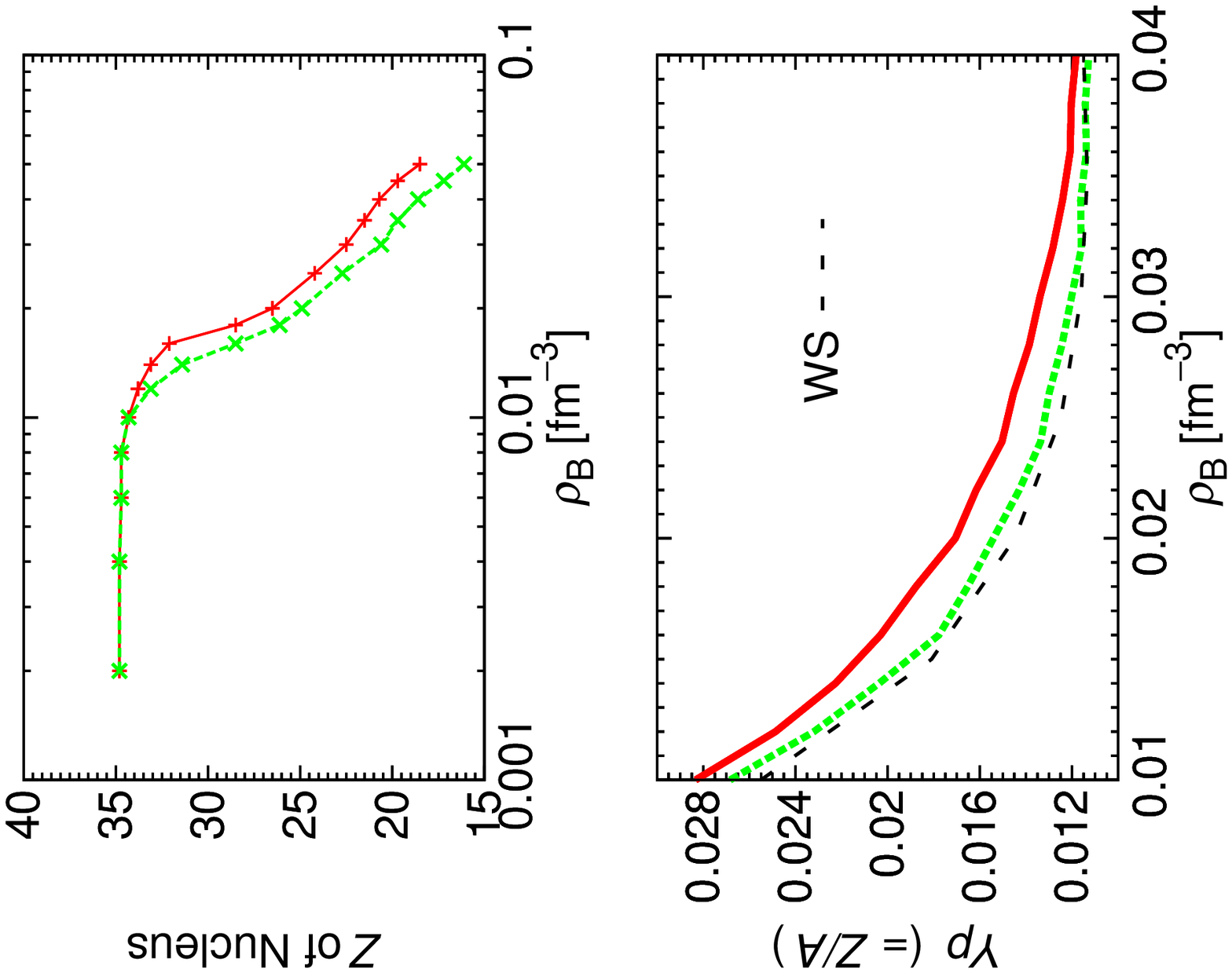}}
  \caption{
  (color online) In the left panel, the density dependence of radii of nuclei,
  lattice constants, and the volume fraction are shown.
  The proton number of nuclei and the proton number-fraction are in the right panel. 
  The lines in each panel indicates fcc by the red line, bcc by the green line in comparison with that in the 
   WS approximation. }
   \label{zyp}
 \end{center}
\end{figure}

As in the case of the fixed proton number-fraction, 
there emerges  a fcc lattice of droplets near the transition density from the droplet phase to another 
in our calculation, 
while it has been regarded to take a bcc lattice in the previous studies \cite{oyamatsu}.  
Almost the same radii and density distribution of droplet, proton number-fraction and proton number in nuclei 
are obtained for both crystalline configurations (see the right panel of Fig.\ \ref{zyp} at $\rho_B<0.01$  fm$^{-3}$). 
The difference between the bcc and fcc lattices may be seen only in the Coulomb energy:  
the Coulomb energy of the fcc lattice is a little higher than that of the bcc lattice. 
However, near the transition density from droplet to rod, 
the radius of droplets and the proton number-fraction are 
different between the bcc and fcc lattices even if 
their baryon-number densities are the same (see Fig.\ \ref{zyp} at $\rho_B\approx 0.02$ fm$^{-3}$): 
the size of the droplet and the proton number-fraction in the fcc lattice is 7.54 fm and 0.016, respectively, 
while those in the bcc lattice are 7.01 fm and 0.014. 
Because the size of droplets and the proton number-fraction are different, 
the Coulomb energy alone is no more the criterion of the ground state. 
We should take into account the size of droplets and the proton number-fraction 
looking for the ground state in a self-consistent way. 
Roughly speaking, the larger the radius of droplets is, the smaller the surface energy is. 
While the Coulomb energy of the fcc lattice is larger than that of the bcc lattice in all the region of the droplet phase, 
the total energy of the fcc lattice is less than that of the bcc lattice by the gain of the surface energy.

Because we cannot take into account the shell effects like Ref.\ \cite{vautherin}, 
the proton number continuously decreases with increase of baryon number density. 
To compare the density dependence of the proton number within the same interaction for baryons, 
our model might be similar to the type B in Ref.\ \cite{iida}. 
In our model, the slope parameter of  symmetry energy $L$ is a little larger than that in Ref.\ \cite{iida},  
so that, in our model, the proton number converges to about 35 in the low-density limit, while about 40 in Ref.\ \cite{iida}. 
This result is consistent with the relationship between $L$ and the proton number in the droplet.

The maximum size of the droplet can be estimated by the Bohr-Wheeler condition 
as $E_{\rm Coul}^{(0)}>2E_{\rm surf}$ \cite{preston}. The "virial" theorem for the pasta structure reads  
$E_{\rm surf}=2E_{\rm Coul}$ where $E_{\rm Coul}^{(0)}$ is 
the Coulomb energy of an isolated nucleus and the Coulomb energy of a nucleus in matter as 
$ E_{\rm Coul} \approx E^{(0)}_{\rm Coul}(1-3u^{1/3}/2)$. 
From these equations, the appearance of non-spherical nucleus in nuclear matter 
has been expected for the volume fraction $u>0.125$. 
However, in our calculation, the structural change from droplet to rod occurs 
around $u\approx 0.1$ (see the value at $\rho_B\approx 0.05$  fm$^{-3}$ in Fig.\ \ref{zyp}). 
This means that considering the non-uniformity of electron is worthwhile for ``pasta'' structures, 
because the relation between the Coulomb energy of a cell and that of a nucleus has been derived 
by using a uniform background electrons and uniform baryon density inside a nucleus. 
The effect of the screening by charged particles, which is properly included in our calculation, 
may be one of the origins of this difference. 
To get the final conclusion of this issue, we should perform another calculation 
with uniformly distributed electrons and confirm the effects of non-uniformity of electron. 
Also the difference of the droplet surface may give another reason: they used 
the CLDM with a sharp surface, while our droplets have a diffuse surface. 
The same discussion may apply  to the QMD calculation 
for the case of the fixed proton number-fraction.

At low density, where the volume fraction of droplet is less than $1\%$,
we can regard nuclei as point particles even if there is a broad distribution of neutron. 
However, near the transition density from droplet to rod, 
it amounts to about $10\%$ and the distance between the nuclei is very close to each other, which 
is one reason of the shape transition. 
In this region, the approximation for each droplet as a point particle is not allowed.

\section{Summary and concluding remarks}

We have numerically explored non-uniform structures and discussed the properties of 
low-density nuclear matter and cold catalyzed matter, using the RMF model under the Thomas-Fermi approximation. 
Without any assumption about the geometric structure, we have carried out fully three-dimensional calculations 
in large cubic cells with the periodic boundary condition.

First, we explored the low-density nuclear matter with the fixed proton number-fraction 
$Y_p=$ 0.5, 0.3, 0.1, which may be relevant to supernova explosions and newly born protoneutron stars. 
With increase of density, which ranges from well below to around the normal nuclear density, 
we have observed the typical pasta structures as a ground state for each density and proton number-fraction. 
The appearance of the pasta structures lowers the energy, 
while the energy differences between various {\it geometrical} structures are very small. 
More exotic structures like ``diamond'', ``dumbbell'', and 
mixtures of two types  of pasta structures appear as metastable states at some transient densities.

Secondly, we extend to the cold catalyzed matter which corresponds to the neutron star crust. 
In this case, with increase of density, which ranges from well below to half of the normal nuclear density, 
we have observed that the ground state of matter shows two types of pasta structures, droplets and rods. 
As for the crystalline configuration of  droplets, near the transition density to rod, 
the fcc lattice is  more favorable than the bcc lattice, which is different from the previous studies. 
We have discussed some reasons of the difference, but more elaborate studies are needed to clarify it.

In this article, the ground state of low-density nuclear matter changes the form 
of crystalline configuration from the bcc to fcc lattice near the half of normal nuclear density within the RMF model. 
This conclusion arises from including the smooth surface of nuclei and the self-consistent calculation of the Coulomb interaction. 
On the other hand, there are several forms of lattice structures like hexagonal closest packing, tetrahedron. 
The hexagonal lattice has the same filling factor with the fcc one. 
There are many possibilities, but is not specific interpretation for the crystalline configuration in the ground state.

In application to neutron star crust, it is interesting and might be important to investigate the shear modulus. 
Using the density distribution obtained by our three-dimensional calculation for cold catalyzed matter (droplet and rod), 
we can estimate the shear modulus, in which the charge screening effect and the finite-size effects are properly taken into account. 
More interestingly, one may directly calculate the breaking strain by considering the large deformation of the lattice.  
These issues will be discussed in a separate paper.

For newly born neutron stars, as in supernova explosions, 
finite temperature and neutrino-trapping effects become important, 
as well as the dynamics of the first order phase transition with formation of the structures. 
It would be also interesting to extend our framework to include these effects.

\section*{Acknowledgement}

We thank H. Sotani for his useful comments.
This work is supported by the project 
{\it ``Research on the Emergence of Hierarchical Structure of Matter by Bridging Particle,
 Nuclear and Astrophysics in Computational Science,'' 2008--2013}, 
of the Grant-In-Aid for Scientific Research on Innovative Areas, MEXT Japan.

\end{document}